\documentstyle{article}
      \makeatletter
      
      \@addtoreset{equation}{section}
      \makeatother

\newcommand{\emp}[1]{{\bf #1}}

\newtheorem{teiri}{\emp{Theorem}}[section]
\newtheorem{kei}{\emp{Corollary}}[section]

\newtheorem{hodai}{\emp{Lemma}}[section]
\newtheorem{teigi}{\emp{Definition}}[section]
\newtheorem{rei}{\emp{Example}}[section]
\newtheorem{chui}{\emp{Remark}}[section]

\newcommand{\QED}{\hspace*{\fill}$\Box$ \\}

\newcommand{\spr}[1]{{\bf #1}}

\newcommand{\vph}{\varphi}

\newcommand{\cA}{{\cal A}}

\newcommand{\cD}{{\cal D}}

\newcommand{\cG}{{\cal G}}

\newcommand{\cM}{{\cal M}}

\newcommand{\cX}{{\cal X}}
\newcommand{\cY}{{\cal Y}}

\newcommand{\sR}{\spr{R}}
\newcommand{\sS}{\spr{S}}

\newcommand{\sX}{\spr{X}}

\newcommand{\sZ}{\spr{Z}}

\newcommand{\ssx}{\spr{x}}

\newcommand{\nth}{\frac{1}{n}}

\newcommand{\noi}{_{n=1}^{\infty}}
\newcommand{\lsn}{\limsup_{n \to \infty}}
\newcommand{\lin}{\liminf_{n \to \infty}}
\newcommand{\lmn}{\lim_{n \to \infty}}

\newcommand{\bteiri}{\begin{teiri}}
\newcommand{\eteiri}{\end{teiri}}
\newcommand{\bkei}{\begin{kei}}
\newcommand{\ekei}{\end{kei}}
\newcommand{\brei}{\begin{rei}}
\newcommand{\erei}{\end{rei}}
\newcommand{\bhodai}{\begin{hodai}}
\newcommand{\ehodai}{\end{hodai}}
\newcommand{\bteigi}{\begin{teigi}}
\newcommand{\eteigi}{\end{teigi}}
\newcommand{\bchui}{\begin{chui}}
\newcommand{\echui}{\end{chui}}
\newcommand{\beq}{\begin{equation}}
\newcommand{\eeq}{\end{equation}}
\newcommand{\beqn}{\begin{eqnarray}}
\newcommand{\eeqn}{\end{eqnarray}}
\newcommand{\beqns}{\begin{eqnarray*}}
\newcommand{\eeqns}{\end{eqnarray*}}

\newcommand{\map}{\vph_n: \cX^n \to \cY^n}
\newcommand{\mapMtoY}{\vph_n: \cM_{M_n} \to \cY^n}
\newcommand{\mapXtoM}{\vph_n: \cX^n \to \cM_{M_n}}

\title{~\\~\\~\\~\\ \bf An Information-Spectrum Approach to Large Deviation Theorems
}
\author{~\\~\\~\\~\\Te Sun HAN\thanks{Te Sun Han is with the Graduate
School of Information Systems, University of Electro-Communications,
Chofugaoka 1-5-1, Chofu, Tokyo 182-8585, Japan. 
E-mail:\ han@is.uec.ac.jp}}

\date{\today}

\begin{document}
\setcounter{page}{0}
\maketitle
\thispagestyle{empty}
\newpage

\pagenumbering{roman}

%
%

\pagenumbering{arabic}
\setcounter{page}{0}
\setcounter{equation}{0}

{\bf Abstract:} \  In this paper we show  some new look at large deviation theorems 
from the viewpoint of the information-spectrum (IS) methods, which has been first exploited  in information theory, and also demonstrate a new basic formula 
for the large deviation rate function in general, which is expressed as a pair of the lower and upper IS rate functions.  In particular, we are interested in 
establishing the general large deviation rate functions that can be derivable as 
 the Fenchel-Legendre transform of the cumulant generating function. The final goal is to
 show a necessary and sufficient condition for the rate function to
  be of Cram\'er-G\"artner-Ellis type.
\newpage

\section{Introduction\label{ss:1}}

The present paper is intended to show a basic new look at  problems in large deviation 
theory. As is well known, in many fields such as information theory, probability theory, computer sciences, 
 communication sciences, and cryptography,  etc.,  we face 
the problem to pertinently evaluate  behaviors of  the tail probability of asymptotic distributions, and it 
is now not only of theoretical importance but also 
of practical interest. 

In such a situation, therefore, it would also be useful to elucidate and look back at
 {\em basic } common  properties  as well as 
 some  common structures underlying the large deviation problem.

For this purpose, we first introduce, as a basic tool for analyzing  the tail probabilities in large deviation,  
 the concept of a  {\em general} source $\sZ=\{Z_n\}\noi$ that is general in the sense that our scope is not only in stationary/ergodic sources but also in non-ergodic and/or nonstationary sources. In order to  enable us to deal with this kind of 
general sources, 
we define a {\em pair} of two kinds of large deviation rate functions, say, a pair of the lower and upper information-spectrum (=IS) rate functions,  basically without any assumptions on its probabilistic 
memory structures. This is because sources
in consideration may be non-ergodic and/or nonstationary. It should be noted here that, by convention, large deviation
rate functions have been supposed to be  assigned with one each source.   Along this line of thought, we 
 try to reveal a general 
skeleton latent in  those basic problems. 
Such an IS approach to various kinds of information-theoretical 
problems has been devised by Han and Verd\'u \cite{han-verdu-approx},
Verd\'u and Han \cite{verdu-han}, and Han \cite{han-book}, etc. 

Several crucial studies on the fundamental large deviation problem are found in 
Varadhan \cite{varadhan} (the integral lemma), Bryc \cite{bryc} (the inverse Varadhan lemma), etc.
In particular, Bryc \cite{bryc} has shown that  a kind of  ``{\em nonlinear}" but
one-to-one transformations (instead of  ``{\em linear}" transformations, say, 
Fenchel-Legendre transformations) is quite useful in order to establish
the {\em good} rate function in general that may not  necessarily be convex, though
the relevant computation of these nonlinear transforms seems to be formidable.

On the other hand, in the present paper, we start with the introduction of 
the notion of lower/upper IS
rate functions to demonstrate
 that the pertinent large deviation rate 
functions (the {\em lower} one and the {\em  upper} one)
 can be reasonably formulated  with the use of these IS
rate functions.

In Section \ref{sec:fuji-dev-A-10s} we prepare the necessary notion 
to establish the {\em pair} of  basic fundamental lower/upper 
large deviation rate functions, where by a ``rate function"
we do not necessarily mean that it is  lower semicontinuous
and all the level sets are closed (cf. Dembo and Zeitouni \cite{demboS}).

In Section \ref{sec:akebono-8} it is shown, as  illustrative applications of the theorems 
stated in Section \ref{sec:fuji-dev-A-10s},  that, for {\em mixed} sources, 
 the lower and upper IS rate functions
coincide
with each other, but they are
  {\em not} convex (cf. Bryc \cite{bryc}, Chen \cite{NoningC}); 
and also that the lower and upper IS rate functions do 
 {\em not} coincide with each other for {\em nonstationary} sources.

In Section \ref{generalization-ex1}, a pleasant  generalization from the real space $\sR\equiv (-\infty, +\infty)$ to the Hausdorff  topological space is pointed out, although this generalization is not used in the subsequent sections. 

In Section \ref{ss:1seoulA}, as a second crucial step, we proceed to elucidate a 
structural correspondence between
the  lower/upper IS
rate functions and the inferior/superior limits of the normalized cumulant 
generating functions  to 
demonstrate the formula for computing the cumulant generating function using the lower/upper IS rate functions, the inverse function of which,
 i.e., the rate function of Cram\'er-G\"artner-Ellis type 
is given in Theorem \ref{teiri:kihonrrs}.  
Here, a  conclusion (Theorem \ref{kei:river-ksks1}) is stated that these are connected equivalently with each other via
the Fenchel-Legendre transformation, under a mild assumption,  if only and if the former 
IS rate functions are closed  and convex, which is called the reduction theorem.

Finally, in Section \ref{setu:proofs} we give the proofs of Theorem \ref{teiri:kihonasBan}
 and Lemma \ref{chui:seoul-han0ajk}.
 %
%

%
%
\section{A General formula for Large Deviations\label{sec:fuji-dev-A-10s}}

\setcounter{equation}{0}
%

%
Let $(Z_1, Z_2, \cdots )$ be any sequence of  random variables
taking values in {\bf R}$\equiv (-\infty, +\infty)$, and call $\sZ = \{Z_n\}\noi$ a {\em general} source.
Here, any probabilistic dependency structures are basically not assumed about 
$\sZ = \{Z_n\}\noi$, where $Z_n$ is  supposed typically to be the arithmetic mean
$\frac{S_n}{n}$ with another underlying ``sum" source $\sS=\{S_n\}\noi$.
We are interested in large deviation behaviors of those $\sZ = \{Z_n\}\noi$.

%
%
%

%
%
%

%
\medskip

Let $\pi_1>\pi_2>\cdots \to 0$ be an  arbitrarily prescribed  sequence. With this 
$\pi_i$ $(i=1,2,\cdots)$
define the shrinking open intervals as follows:
\beq\label{seoul-sita12}
\Phi_{i}(R) \equiv (R-\pi_i, R+\pi_i)\quad (i =1,2, \cdots),
\eeq
and also define 
as follows:
\bteigi\label{teigi:extra-1}
\beqn\label{seoul-sita13}
\underline{H}_i(R) & = & \lin \frac{1}{n}\log\frac{1}{\Pr\{Z_n \in \Phi_i(R)\}},
\label{seoul-sita14}
\\
\overline{H}_i(R) & = & \lsn \frac{1}{n}\log\frac{1}{\Pr\{Z_n \in \Phi_i(R)\}},
\label{seoul-sita15}
\eeqn
\beqn
\underline{H}(R) & = & \lim_{i\to\infty}\underline{H}_i(R),\label{seoul-sita16}
\\
\overline{H}(R) & = & \lim_{i\to\infty}\overline{H}_i(R).
\label{seoul-sita17}
\eeqn
\eteigi \QED
Clearly, $\underline{H}_i(R)$ and $ \overline{H}_i(R)$ are increasing functions in $i$.
It is also obvious that $\underline{H}_i(R) \le \overline{H}_i(R)$  $(i=1,2, \cdots)$  and $\underline{H}(R) \le \overline{H}(R)$.  We call these $\underline{H}(R), \overline{H}(R)$ the lower/upper IS rate functions of $\sZ = \{Z_n\}\noi$.
It should  also be  remarked that $\underline{H}(R), \overline{H}(R)$ do not depend on
the choice of the sequence $\pi_1>\pi_2>\cdots \to 0$.

Throughout in this paper all relevant  quantities such as  $\underline{H}(R), \overline{H}(R)$ are allowed to take  values  $\pm\infty$.
%
%
%
\medskip

In order to establish fundamental  formulas for ``Large Deviation Principle" (=LDP), 
 we need here the following notion.
\bteigi{\rm ($E$-tight)}\label{ISITA-S1}
Let $\sZ = \{Z_n\}\noi$ be a general source.
If
\beq\label{eq:cosmosq-AB}
\lim_{K\to\infty}\lsn\nth\log \Pr\{|Z_n|>K\} = -\infty
\eeq
holds, then we say that the source $\sZ=$ $\{Z_n\}\noi$ is
 exponentially tight (abbreviated as $E$-tight;\ cf. Dembo and Zeitouni \cite{demboS}).
\eteigi

Here, we have the following   fundamental  theorem for LDP,
although it has a rather conventional  form. This theorem can also be regarded as forming a pair with Theorem \ref{teiri:spec-lower}  below.
\bteiri\label{teiri:spec-upper}
If a general source $\sZ=\{Z_n\}\noi$ is $E$-tight, 
then for any measurable set
 $\Gamma$ it holds that 
\beq\label{eqk:osamu-41}
-\inf_{R\in \Gamma^{\circ}}\underline{H}(R)\le 
\lsn \nth\log \Pr\{Z_n \in \Gamma\}
 \le -\inf_{R\in \overline{\Gamma}}\underline{H}(R),
\eeq
where $\Gamma^{\circ}$, $\overline{\Gamma}$ are the interior  and 
the closure of $\Gamma$, respectively.
\eteiri

\bchui\label{chui:kenkou-3}
As will be seen in the proof below, the lower bound in
inequality (\ref{eqk:osamu-41}):
\beq\label{eqk:kenkou-1}
 -\inf_{R\in \Gamma^{\circ}}\underline{H}(R)
\le 
\lsn \nth\log \Pr\{Z_n \in \Gamma\}
\eeq
holds without the $E$-tightness  assumption. On the other hand, 
as for another type of upper bound for (\ref{eqk:osamu-41}), 
see Remark \ref{chui:ranranQRS2}
in Section \ref{ss:1seoulA} ( also see Corollary \ref{kei-hansi-God} below).
%
\echui

\bchui\label{chui:akebono-C}
The $E$-tightness  (Definition \ref{ISITA-S1}) of the source has the following meaning for the lower
IS rate function $\underline{H}(R)$:
let $K>0$ be an arbitrarily large number, then for any $R$ such that $|R|>K$ it holds that 
\[
\lsn \nth\log \Pr\{|Z_n|>K\} \ge -\underline{H}(R)
\]
in view of the definition of  $\underline{H}(R)$. Hence,
\[
\lsn \nth\log \Pr\{|Z_n|>K\} \ge -\inf_{|R|>K}\underline{H}(R).
\]
Therefore, from the $E$-tightness (\ref{eq:cosmosq-AB}) we have.
\[
\lim_{K\to\infty}\inf_{|R|>K}\underline{H}(R) =+\infty,
\]
that is, 
\beq\label{eqk:kruger-4s}
\liminf_{R\to+\infty}\underline{H}(R) =+\infty,
\quad \liminf_{R\to-\infty}\underline{H}(R) =+\infty.
\eeq
However，(\ref{eqk:kruger-4s}) does not necessarily imply the $E$-tightness.
For example, if we consider the source $\sZ=\{Z_n\}\noi$ such that $\Pr\{Z_n=n\}=$ $\Pr\{Z_n=-n\}=$ $\frac{1}{2}$
$(\forall n=1,2,\cdots)$, this leads to 
$\underline{H}(R) =\overline{H}(R) =+\infty$ $(\forall R\in\sR)$, and so in this case  (\ref{eqk:kruger-4s}) holds but (\ref{eq:cosmosq-AB}) does not hold．
\QED
\echui

\noindent
{\em Proof of Theorem \ref{teiri:spec-upper}}:
The proof is quite elementary.
 It is enough to show 
\beq\label{eq:kenkou-101}
\lsn \nth\log \Pr\{Z_n \in \Gamma\}
 \le -\inf_{R\in \overline{\Gamma}}\underline{H}(R)
\eeq
and (\ref{eqk:kenkou-1}).

\medskip

\noindent 
a)\ First we show (\ref{eqk:kenkou-1}). For notational simplicity we use here the notation 
$\Gamma_{\delta}(R) $ $\equiv (R - \delta，R+ \delta)$ (cf. (\ref{seoul-sita12})).
Then, for any small number $\delta >0$ it holds that 
\beq\label{eqw:tepo-2}
\Gamma_{\delta}(R)\supset \Phi_{i}(R)
\quad (\forall i \ge  i_0( \delta)).
\eeq
On the other hand, from the definition (\ref{seoul-sita16}) of 
$\underline{H}(R)$ it follows that
\beq\label{eq:kenkou-102}
\lsn\nth\log \Pr\{Z_n \in \Phi_i(R)\}
 \ge -(\underline{H}(R)+\gamma)
\quad ( \forall i \ge i_0(R)),
\eeq
where  $\gamma>0$ is an arbitrary small number.
Therefore, there exists a sequence $n_1<n_2<\cdots \to\infty$ 
(dependent on $R$) such that
\beq\label{eqw:tepo-4}
\frac{1}{n_k}\log \Pr\{Z_{n_k} \in \Phi_i(R)\}
 \ge -(\underline{H}(R)+2\gamma)
\quad (\forall i \ge i_0(R)).
\eeq
Thus, in view of (\ref{eqw:tepo-2}) and (\ref{eqw:tepo-4}),
\[
\frac{1}{n_k}\log \Pr\{Z_{n_k} \in \Gamma_{\delta}(R)\}
 \ge -(\underline{H}(R)+2\gamma).
\]
Hence,
\beqns
\lefteqn{\lsn\nth\log \Pr\{Z_{n} \in \Gamma_{\delta}(R)\}}\nonumber\\
& \ge & 
\limsup_{k\to\infty}\frac{1}{n_k}
\log \Pr\{Z_{n_k} \in \Gamma_{\delta}(R)\}\\
&  \ge & -(\underline{H}(R)+2\gamma).
\eeqns
As $\gamma>0$ is arbitrary, we have 
\beq\label{eqw:tepo-5}
\lsn\nth\log \Pr\{Z_n \in \Gamma_{\delta}(R)\}
 \ge -\underline{H}(R)\quad (\forall \delta >0).
\eeq
Here for any $R \in \Gamma^{\circ}$ let $\delta >0$ be small enough to satisfy $\Gamma_{\delta}(R) \subset \Gamma^{\circ}$. Then, by means of (\ref{eqw:tepo-5}),\beqns
\lsn \nth\log \Pr\{Z_n \in \Gamma\} & \ge &
 \lsn \nth\log \Pr\{Z_n \in \Gamma^{\circ}\}\\
&\ge & \lsn \nth\log \Pr\{Z_n \in \Gamma_{\delta}(R)\} \\
&\ge &  -\underline{H}(R).
\eeqns
As $R$ is an arbitrary internal point of $\Gamma^{\circ}$, 
we conclude that
\[
\lsn \nth\log \Pr\{Z_n \in \Gamma\} \ge 
-\inf_{R\in \Gamma^{\circ}}\underline{H}(R).
\]

\medskip

\noindent
b)\ Next we will show (\ref{eq:kenkou-101}).
It follows from the definition  (\ref{seoul-sita16}) of 
$\underline{H}(R)$ that
\beqn\label{eqk:water-31}
\nth\log \Pr\{Z_n \in \Phi_{i}(R)\} \le -(\underline{H}(R)-2\gamma)
& &  ( \forall n \ge n_0(R, i);\ \forall i\ge  i_0(R)),
\nonumber\\
& &
\eeqn
where 
 $\gamma>0$ is an arbitrarily small number.
 For any constant $K>0$ set 
$\overline{\Gamma}_K\equiv$ $\overline{\Gamma} \cap [-K, K]$
 and consider an arbitrary 
$R\in \overline{\Gamma}_K$, then (\ref{eqk:water-31})
with $ i=i_0\equiv i_0(R)$ reduces to 
\beq\label{eqk:boston-309-1}
\nth\log \Pr\{Z_n\in \Phi_0(R)\}
 \le
-(\underline{H}(R)-2\gamma)\quad  (\forall n \ge n_0(R, i_0); \forall R \in \overline{\Gamma}_K),
\eeq
where we have put $\Phi_0(R) = \Phi_{i_0}(R)$. It is evident that $\Phi_0(R)$ is an open set. 
Now we notice that $\overline{\Gamma}_K $ is a bounded closed set and hence satisfies the compactness (owing to Heine-Borel theorem). As a consequence, we can choose a finite number of points $R_1, R_2, \cdots, R_{m_K}$ in  $\overline{\Gamma}_K$ so that
\[
 \overline{\Gamma}_K \subset 
\bigcup_{l=1}^{m_K}\Phi_0(R_l).
\]
Therefore, for
\[
\forall n \ge \max\left(n_0(R_1, i_0), n_0(R_2,i_0), \cdots, n_0(R_{m_K}, i_0)\right),
\]
we have
\beqns
\lefteqn{\nth\log \Pr\{Z_n\in \overline{\Gamma}_K\}}\\
& \le & \nth\log \left(\sum_{l=1}^{m_K}
\Pr\{Z_n\in \Phi_0(R_l)\}\right)\\
& \le & 
\nth\log\left(\max_{1\le l \le m_K}\Pr\{Z_n\in \Phi_0(R_l)\}\right)
+\nth \log m_K.
\eeqns
Hence, by (\ref{eqk:boston-309-1}) we have 
\beqn\label{eqk:boston-309-2}
\lefteqn{\nth\log \Pr\{Z_n\in \overline{\Gamma}_K\}}\nonumber\\
& \le & -\min_{1\le l \le m_K}\underline{H}(R_l) +\nth \log m_K+2\gamma
\nonumber\\
&\le & -\inf_{R\in \overline{\Gamma}_K}\underline{H}(R)  +\nth \log m_K+2\gamma,\nonumber
\eeqn
from which it follows that
\[
\lsn \nth\log \Pr\{Z_n\in \overline{\Gamma}_K\}
\le   -\inf_{R\in \overline{\Gamma}_K}\underline{H}(R)+2\gamma.
\]
Then, as  $\gamma>0$ is arbitrary, 
\beq\label{eqk:boston-309-3}
\lsn \nth\log \Pr\{Z_n\in \overline{\Gamma}_K\}
\le  -\inf_{R\in \overline{\Gamma}_K}\underline{H}(R).
\eeq
On the other hand,  the assumed $E$-tightness condition  
of $\sZ=\{Z_n\}\noi$ implies that for any large 
$L>0$,
\beq\label{eqk:boston-310-1}
\lsn\nth\log \Pr\{|Z_n|>K\}
\le -L \quad (\forall K\ge  K_0(L)).
\eeq
Hence, combining (\ref{eqk:boston-309-3}) and
(\ref{eqk:boston-310-1}) results in 
\beqn\label{eqk:cosmosq-8-119-B}
\lefteqn{\lsn\nth\log \Pr\{Z_n\in \overline{\Gamma}\}}\nonumber\\
& \le & \lsn\nth\log \left(\Pr\{Z_n\in \overline{\Gamma}_K\}+
\Pr\{|Z_n|>K\}\right)\nonumber\\
&\le & \lsn \left(\nth\log \max\left(\Pr\{Z_n\in \overline{\Gamma}_K\},
\Pr\{|Z_n|>K\}\right)+\nth\log 2\right)\nonumber\\
&= & \lsn \left( \max\left(\nth\log\Pr\{Z_n\in \overline{\Gamma}_K\},
\nth\log\Pr\{|Z_n|>K\}\right)\right)\nonumber\\
&= & \max\left(\lsn\nth\log \Pr\{Z_n\in \overline{\Gamma}_K\},
\lsn\nth\log \Pr\{|Z_n|>K\}\right)\nonumber\\
&\le &
 \max\left(-\inf_{R\in \overline{\Gamma}_K}\underline{H}(R), 
 -L\right)\nonumber\\
&=& -\min\left(\inf_{R\in \overline{\Gamma}_K}\underline{H}(R),
 L\right)\nonumber\\
 & \le &  -\min\left(\inf_{R\in \overline{\Gamma}}\underline{H}(R),
 L\right).
\eeqn
We notice here that $L>0$ is arbitrarily large, so letting 
$L\to\infty$ we have%
\beqns
\lsn \nth\log \Pr\{Z_n\in \Gamma\}
& \le & \lsn \nth\log \Pr\{Z_n\in \overline{\Gamma}\}\\
& \le &  -\inf_{R\in \overline{\Gamma}}\underline{H}(R),
\eeqns
thus completing the proof of (\ref{eq:kenkou-101}). \QED

\medskip

Let us now proceed to show the second fundamental theorem on large deviation.
To do so, we need  the following notion.
\bteigi{\rm ($\sigma$-convergence)}\label{ISITA-S2}
Let $f_n(R) \ (n=1,2,\cdots)$ be a sequence of functions of $R$ on {\bf R}.
  If for any bounded closed subset  $\cD$ of {\bf R} and  for any $\gamma>0$ 
 there exists a sequence $n_1<n_2<\cdots \to +\infty$ (independent of $R$ $\in$
 $\cD$)  such that 
\beq\label{eq:cosmosq-B}
f_{n_k}(R) \ge \lsn f_n(R)  - \gamma \quad (\forall k \ge k_0(R, \gamma); \forall R \in \cD ),
\eeq
then we say that $\{f_n(R)\}\noi$ is $\sigma$-convergent.
\eteigi
\bchui\label{sita-seoull-1}
It is easy to check that $\{f_n(R)\}\noi$ is  $\sigma$-convergent if $\{f_n(R)\}\noi$  converges on {\bf R}. As for the IS meaning of Definition 
\ref{ISITA-S2}, refer to Definition \ref{ISITA-S2w} below. \QED
\echui

\bteigi{\rm ($\sigma$-convergent)}\label{ISITA-S2w}
Given a general source  $\sZ=$ $\{Z_n\}\noi$, set
\beq\label{eq:han1}
f_n(R) \equiv \frac{1}{n}\log \frac{1}{\Pr\{Z_n \in \Phi_i(R)\}}\quad (\mbox{with any fixed $i$}),
\eeq
where $\Phi_i(R)$ is defined as  in  ( \ref{seoul-sita12}).
If  $\{f_n(R)\}\noi$ is $\sigma$-convergent in the sense of Definition \ref{ISITA-S2}, then we say that
the source $\sZ=$ $\{Z_n\}\noi$ is 
$\sigma$-convergent .\QED
\eteigi

\bchui\label{chui:Seoul1}
It is not difficult to check that $\overline{H}(R) = \underline{H}(R)$ 
$(\forall R \in \sR)$
is a sufficient condition for the $\sigma$-convergence of $\sZ=$ $\{Z_n\}\noi$. 
\QED
\echui

With these definitions,  we have the following second fundamental theorem,
which can be regarded as providing the pair with Theorem \ref{teiri:spec-upper}.

\bteiri\label{teiri:spec-lower}
If a general source $\sZ=\{Z_n\}\noi$ is $E$-tight and $\sigma$-convergent, then for any measurable set $\Gamma$ it holds that\beq\label{eq:dragon-11}
 -\inf_{R\in \Gamma^{\circ}}\overline{H}(R)
\le 
\lin \nth\log \Pr\{Z_n \in \Gamma\}
\le
 -\inf_{R\in \overline{\Gamma}}\overline{H}(R),
\eeq
where $\Gamma^{\circ}$, $\overline{\Gamma}$ are the interior  and 
the closure of $\Gamma$, respectively．\QED
\eteiri

\bchui\label{chui:kenkou-1}
As will be seen from the proof below, the lower bound in inequality (\ref{eq:dragon-11}):
\beq\label{eq:kenkou-1}
 -\inf_{R\in \Gamma^{\circ}}\overline{H}(R)
\le 
\lin \nth\log \Pr\{Z_n \in \Gamma\}
\eeq
holds without the assumptions of Theorem \ref{teiri:spec-lower}．
\QED
\echui

\noindent
{\em Proof of Theorem \ref{teiri:spec-lower}:}\ 
The proof is quite elementary.
It suffices to show
\beq\label{eq:kenkou-2}
 \lin \nth\log \Pr\{Z_n \in \Gamma\}
\le  -\inf_{R\in \overline{\Gamma}}\overline{H}(R)
\eeq
and (\ref{eq:kenkou-1}).

\medskip

\noindent
a)\ First we show (\ref{eq:kenkou-1}).
Although the proof of this part basically parallels that of part a) of Theorem \ref{teiri:spec-upper}
 with due modifications, we write it again for the reader's convenience.
 Here too, we use  the notation that
$\Gamma_{\delta}(R) $ $\equiv (R - \delta，R+ \delta)$
($\delta>0$). Then, we have
\beq\label{eqw:tepo-2Z}
\Gamma_{\delta}(R)\supset \Phi_{i}(R)
\quad (\forall i \ge  i_0( \delta)).
\eeq
Moreover, by the definition (\ref{seoul-sita17}) of 
$\overline{H}(R)$,
\beq\label{eq:tepo-3}
\lin\nth\log \Pr\{Z_n \in \Phi_{i}(R)\}
 \ge -(\overline{H}(R)+\gamma)
\quad ( \forall i \ge  i_0(R)),
\eeq
where  $\gamma>0$ is an arbitrarily small number.
Hence, 
\beq\label{eq:tepo-4}
\nth\log \Pr\{Z_n \in\Phi_{i}(R)\}
 \ge -(\overline{H}(R)+2\gamma)
\quad ( \forall n\ge  n_0(R, i);\ \forall i \ge  i_0(R)).
\eeq
Therefore，by (\ref{eqw:tepo-2Z}) and (\ref{eq:tepo-4}),
\[
\nth\log \Pr\{Z_n \in \Gamma_{\delta}(R)\}
 \ge -(\overline{H}(R)+2\gamma)
\quad ( \forall n\ge  n_0(R, \delta)).
\]
Hence,
\[
\lin\nth\log \Pr\{Z_n \in \Gamma_{\delta}(R)\}
 \ge -(\overline{H}(R)+2\gamma).
\]
As $\gamma>0$ is arbitrary, we have
\beq\label{eq:tepo-5}
\lin\nth\log \Pr\{Z_n \in \Gamma_{\delta}(R)\}
 \ge -\overline{H}(R).
\eeq
Now, for any $R\in \Gamma^{\circ}$ we can choose a small $\delta>0$ so that 
$\Gamma_{\delta}(R) \subset \Gamma^{\circ}$. Then, by (\ref{eq:tepo-5}),
\beqns
\lin \nth\log \Pr\{Z_n \in \Gamma\} & \ge &
 \lin \nth\log \Pr\{Z_n \in \Gamma^{\circ}\}\\
&\ge & \lin \nth\log \Pr\{Z_n \in \Gamma_{\delta}(R)\} \\
&\ge &  -\overline{H}(R).
\eeqns
Since $R\in \Gamma^{\circ}$ is arbitrary, we conclude that
\[
\lin \nth\log \Pr\{Z_n \in \Gamma\} \ge 
-\inf_{R\in \Gamma^{\circ}}\overline{H}(R), 
\]
which implies (\ref{eq:kenkou-1}).

\medskip

\noindent
b)\  Next we show (\ref{eq:kenkou-2}).
With an arbitraily large $K>0$  we set 
$\overline{\Gamma}_K\equiv$ $\overline{\Gamma} \cap [-K, K]$.
 By the definition (\ref{seoul-sita17}) of $\overline{H}(R)$,
\beqn\label{eq:kenkou-4}
\overline{H}_{i}(R) &\equiv & \lsn \nth\log\frac{1}{\Pr\left\{Z_n\in  \Phi_{i}(R)\right\}}\nonumber\\
& \ge &   \overline{H}(R) -\gamma
\quad (\forall i\ge  i_0(R)).
\eeqn
 Then, by the assumed $\sigma$-convergence  of $\sZ=\{Z_n\}\noi$,
  there exists a sequence $n_1< n_2< \cdots \to +\infty$ (independent of $R$ $\in $
  $\overline{\Gamma}_K$ ) such that
\beqn\label{eq:kenkou-5}
\frac{1}{n_k}\log \frac{1}{\Pr\{Z_{n_k} \in \Phi_{i}(R)\}}
 & \ge  & \overline{H}_{i}(R) -2\gamma\nonumber\\
 \quad
& & (\forall k\ge k_0(R, i);\  \forall i  \ge  i_0(R); \  \forall R \in \overline{\Gamma}_K).
\eeqn
As a consequence, combining (\ref{eq:kenkou-4}) and (\ref{eq:kenkou-5})
yields
\beqn\label{eq:kenkou-6}
\Pr\{Z_{n_k} \in \Phi_{i}(R)\}
 &\le &
\exp[-n_k(\overline{H}(R) -3\gamma)]
\nonumber\\
& & \quad\quad\quad (\forall k\ge k_0(R, i);\  \forall i \ge  i_0(R); \forall R \in \overline{\Gamma}_K).
\eeqn
Consider the special case of (\ref{eq:kenkou-6})  with $i=i_0\equiv i_0(R),$
and put $\Phi_0(R) =\Phi_{i_0}(R)$. Then, (\ref{eq:kenkou-6}) reduces to 
\beqn\label{eq:boston-309-1}
\lefteqn{\frac{1}{n_k}\log \Pr\{Z_{n_k}\in \Phi_0(R)\}}\nonumber\\
 & \le &
-(\overline{H}(R)-3\gamma)
\quad (\forall k \ge k_0(R);\ \forall R\in \overline{\Gamma}_K).
\eeqn
We note here that $\overline{\Gamma}_K$ is a bounded closed set and hence  is
compact (owing to Heine-Borel theorem).
Therefore, there exists a finite number of 
$R_1, R_2,\cdots, R_{m_K} \in \overline{\Gamma}_K $ such that
\[
 \overline{\Gamma}_K \subset 
\bigcup_{l=1}^{m_K}\Phi_0(R_l).
\]
Thus, 
\beqns
\lefteqn{\frac{1}{n_k}\log \Pr\{Z_{n_k}\in \overline{\Gamma}_K\}}\\
& \le & \frac{1}{n_k}\log \left(\sum_{l=1}^{m_K}
\Pr\{Z_{n_k}\in \Phi_0(R_l)\}\right)\\
& \le & 
\frac{1}{n_k}\log
\left(\max_{1\le l \le m_K}\Pr\{Z_{n_k}\in \Phi_0(R_l)\}\right)
+\frac{1}{n_k} \log m_K,
\eeqns
from which together with (\ref{eq:boston-309-1}) it follows that
\beqn\label{eq:boston-309-2}
\lefteqn{\frac{1}{n_k}\log \Pr\{Z_{n_k}\in \overline{\Gamma}_K\}}
\nonumber\\
& \le & -\min_{1\le l \le m_K}\overline{H}(R_l) +\frac{1}{n_k} \log m_K+3\gamma
\nonumber\\
& \le & -\inf_{R\in \overline{\Gamma}_K}\overline{H}(R) +\frac{1}{n_k} \log m_K+3\gamma.
\nonumber
\eeqn
Hence, 
\beqns
\lin \nth\log \Pr\{Z_n\in \overline{\Gamma}_K\}
& \le  &
\liminf_{k\to\infty}\frac{1}{n_k}
\log \Pr\{Z_{n_k}\in \overline{\Gamma}_K\}\\
& \le & -\inf_{R\in \overline{\Gamma}_K}\overline{H}(R)+3\gamma.
\eeqns
As $\gamma>0$ is arbitrary, we have
\beq\label{eq:boston-309-3}
\lin \nth\log \Pr\{Z_n\in \overline{\Gamma}_K\}
\le  -\inf_{R\in \overline{\Gamma}_K}\overline{H}(R).
\eeq
On the other hand, 
by the assumed $E$-tightness condition of the source $\sZ=\{Z_n\}\noi$,
for any large $L>0$, 
\beq\label{eq:boston-310-1}
\lsn\nth\log \Pr\{|Z_n|>K\}
\le -L \quad (\forall K\ge  K_0(L)).
\eeq
Then，from (\ref{eq:boston-309-3}) and
(\ref{eq:boston-310-1}),
\beqn\label{eq:cosmosq-8-119-B}
\lefteqn{\lin\nth\log \Pr\{Z_n\in \overline{\Gamma}\}}\nonumber\\
& \le & \lin\nth\log \left(\Pr\{Z_n\in \overline{\Gamma}_K\}+
\Pr\{|Z_n|>K\}\right)\nonumber\\
&\le & \lin \left(\nth\log \max\left(\Pr\{Z_n\in \overline{\Gamma}_K\},
\Pr\{|Z_n|>K\}\right)+\nth\log 2\right)\nonumber\\
&= & \lin \left( \max\left(\nth\log\Pr\{Z_n\in \overline{\Gamma}_K\},
\nth\log\Pr\{|Z_n|>K\}\right)\right)\nonumber\\
&\le & \max\left(\lin\nth\log \Pr\{Z_n\in \overline{\Gamma}_K\},
\lsn\nth\log \Pr\{|Z_n|>K\}\right)\nonumber\\
&\le &
 \max\left(-\inf_{R\in \overline{\Gamma}_K}\overline{H}(R), -L\right)\nonumber\\
&=& -\min\left(\inf_{R\in \overline{\Gamma}_K}\overline{H}(R),
L\right)\nonumber\\
&\le & -\min\left(\inf_{R  \in \overline{\Gamma}}\overline{H}(R),
L\right).
\eeqn
We notice here that $L>0$ is arbitrarily large, so letting $L\to \infty$ we have
\beqns
\lin \nth\log \Pr\{Z_n\in \Gamma\}
& \le & \lin \nth\log \Pr\{Z_n\in \overline{\Gamma}\}\\
& \le &  -\inf_{R\in \overline{\Gamma}}\overline{H}(R),
\eeqns
which implies (\ref{eq:kenkou-2})．\QED

\medskip

So far we have demonstrated two fundamental formulas for large deviation
( Theorem \ref{teiri:spec-upper} and
Theorem \ref{teiri:spec-lower}), which are quite basic  from the viewpoint of 
information-spectra. It should be noted here that  
\[
\lsn \nth\log \Pr\{Z_n \in \Gamma\}
\quad \mbox{and}\quad  \lin \nth\log \Pr\{Z_n \in \Gamma\}
\]
 are both given their own  lower and upper bounds, respectively, while in usual large deviation theorems 
$\lsn \nth\log \Pr\{Z_n \in \Gamma\}
$ and \ $\lin \nth\log \Pr\{Z_n \in \Gamma\}$ are altogether given a pair of  lower  and  upper bounds.

\medskip

From this  conventional standpoint,
 we can specialize Theorem \ref{teiri:spec-upper}, 
Theorem \ref{teiri:spec-lower} along with Remark \ref{chui:kenkou-3},
Remark \ref{chui:kenkou-1} to the following 
two corollaries ({\em full} LDP and {\em weak} LDP):
\bkei\label{sing-Seoul1}{\rm (Full LDP)}
Let a source $\sZ=$ $\{Z_n\}\noi$ be $E$-tight, and suppose that
the IS rate functions have the limit $H(R)$, i.e., 
 $H(R)\equiv \underline{H}(R)=\overline{H}(R)$  ($ \forall R \in \sR$). Then, for any measurable set
$\Gamma$,
\beqn\label{Daveq1}
-\inf_{R\in \Gamma^{\circ}}H(R)
&\le &
\lin \nth\log \Pr\{Z_n \in \Gamma\}\nonumber\\
  &\le &\lsn \nth\log \Pr\{Z_n \in \Gamma\}
  \le -\inf_{R\in \overline{\Gamma}}H(R).\eeqn
\ekei
\bchui\label{chui:nemuiq}
The $E$-tightness condition in Corollary \ref{sing-Seoul1} is actully needed
in order to make the full LDP hold..
For example, let us consider the source $\sZ=\{Z_n\}\noi$ as was shown in 
 Remark \ref{chui:akebono-C}.
 Then,
\[
\underline{H}(R) = \overline{H}(R)=+\infty\quad (\forall R \in \sR),
\]
\[
\lmn \nth\log \Pr\{Z_n \in \Gamma\} =0 \quad(\Gamma=\sR), 
\]
which obviously contradicts (\ref{Daveq1}). \QED
\echui
\bkei\label{kei-hansi-God}{\rm (Weak LDP: Dembo and Zeitouni \cite{demboS})}
\mbox{}
Suppose that  the IS rate functions have the limit $H(R)$, i.e., 
 $H(R)\equiv \underline{H}(R)=\overline{H}(R)$  ($ \forall R \in \sR$).
  Then, the following
{\em weak} LDP holds:

\medskip

\noindent
\quad 1)\ For any set $\Gamma$,
\beq\label{eq:lanlanbc1}
-\inf_{R\in \Gamma^{\circ}}H(R)
\le 
\lin \nth\log \Pr\{Z_n \in \Gamma\};
\eeq

\noindent
\quad 2)\ For any compact set $\Gamma$,
\beq\label{eq:lanlanbc2}
\lsn \nth\log \Pr\{Z_n \in \Gamma\}
  \le -\inf_{R\in \Gamma}H(R).
\eeq

\ekei
\section{ Examples\label{sec:akebono-8}}
\setcounter{equation}{0}

In this section, as  illustrative examples of the fundamental formulas as shown in the 
foregoing section,  let us consider the large deviation behavior of {\em mixed} sources  and/or {\em nonstationary sources} (cf. Han \cite{han-book}).

\bigskip

\noindent
{\em A.\ Mixed sources:}
\medskip

First, let
\beq\label{eq:donuts-1}
\sX_1 = (X_{1,1}, X_{1,2}, \cdots),
\eeq
\beq\label{eq:donuts-2}
\sX_2 = (X_{2,1}, X_{2,2}, \cdots)
\eeq
be two stationary memoryless Gaussian sources with values in {\bf R}, and
let the probability distributions of
\[
X^n_1 =(X_{1,1}, X_{1,2}, \cdots, X_{1,n}),
\]
\[
X^n_2 =(X_{2,1}, X_{2,2}, \cdots, X_{2,n})
\]
be denoted by $P_{X^n_1}(\cdot)$, $P_{X^n_2}(\cdot)$, respectively. 
Moreover, let  the {\em mixed} source of $\sX_1$ and $ \sX_2$:  
\beq\label{eq:donuts-4}
\sX = (X_{1}, X_{2}, \cdots)
\quad
(X^n=(X_{1}, X_{2}, \cdots, X_n))
\eeq
be defined as the source with the probability distribution:
\beq\label{eq:donuts-3}
P_{X^n}(d\ssx) =\alpha_1P_{X^n_1}(d\ssx)+\alpha_2P_{X^n_2}(d\ssx)
\quad (n=1,2,\cdots;\ \ssx \in \sR^n),
\eeq
where $\alpha_1>0, \alpha_2>0$ are constants such that
$\alpha_1+\alpha_2=1$. The mixed source $\sX$ thus defined is 
not memoryless but stationary.

Setting
\beq\label{eq:donuts-5}
Z_{n} = \frac{1}{n}\sum_{i=1}^nX_{i},
\eeq
we are interested in  the large deviation behavior of $\sZ=\{Z_n\}\noi$.
Put
\beq\label{eq:donuts-6}
Z_{1,n} = \frac{1}{n}\sum_{i=1}^nX_{1,i},
\eeq
\beq\label{eq:donuts-7}
Z_{2,n }= \frac{1}{n}\sum_{i=1}^nX_{2,i}
\eeq
and with a fixed number $R_0$ let $\Phi_i(R_0)$ be such as defined in
(\ref{seoul-sita12}) in Section \ref{sec:fuji-dev-A-10s}．
First, by Cram\'er's theorem for the arithmetic mean of a stationary memoryless source
(cf. Dembo and Zeitouni \cite{demboS}), we have
\beqn\label{eq:donuts-9}
-\inf_{R \in \Phi_i(R_0)}I_1(R) & \le &
\lin\nth\log\Pr\{Z_{1,n} \in \Phi_i(R_0)\} \nonumber\\
& \le &
\lsn\nth\log\Pr\{Z_{1,n} \in \Phi_i(R_0)\} \le
-\inf_{R \in \overline{\Phi_i(R_0)}}I_1(R), \nonumber\\
& &
\eeqn
where $I_1(R)$ is the large deviation rate function of the stationary memoryless 
source $\sZ_1 = \{Z_{1,n}\}\noi$
that is defined as 
\beq\label{eq:donuts-10}
I_1(R) = \sup_{\theta}(\theta R -\varphi_1(\theta))
\eeq
in terms of the cumulant generating function 
$\varphi_1(\theta)\equiv$ $\log\mbox{E}(e^{\theta X_{1,1}}) $ of $X_{1,1}$.
Since 
$I_1(R)$ is a closed convex function 
( and hence a continuous function) of $R$ and 
$\Phi_i(R_0)$ is a nonempty  open interval, we have
\[
\inf_{R \in \Phi_i(R_0)}I_1(R)=
\inf_{R \in \overline{\Phi_i(R_0)}}I_1(R).
\]
Then, by virtue of (\ref{eq:donuts-9}), we see that 
$\nth\log\Pr\{Z_{1,n} \in \Phi_i(R_0)\} $ has the limit (as $n\to\infty$), and
(\ref{eq:donuts-9}) is written as 
\beq\label{eq:donuts-11A}
\lmn\nth\log\Pr\{Z_{1,n} \in  \Phi_i(R_0)\} =
-\inf_{R \in  \Phi_i(R_0)}I_1(R).
\eeq
In an analogous manner, we have
\beq\label{eq:donuts-12B}
\lmn\nth\log\Pr\{Z_{2,n} \in  \Phi_i(R_0)\} =
-\inf_{R \in \Phi_i(R_0)}I_2(R), 
\eeq
 where $I_2(R)$ is the large deviation rate function of the stationary memoryless 
source $\sZ_2 = \{Z_{2,n}\}\noi$
that is defined as 
\beq\label{eq:donuts-10-23}
I_2(R) = \sup_{\theta}(\theta R -\varphi_2(\theta))
\eeq
in terms of the cumulant generating function 
$\varphi_2(\theta)\equiv$ $\log\mbox{E}(e^{\theta X_{2,1}}) $  of $X_{2,1}$.
On the other hand, 
in view of (\ref{eq:donuts-3}) we see that
\[
\Pr\{Z_{n} \in \Phi_i(R_0)\}=
\alpha_1\Pr\{Z_{1,n} \in \Phi_i(R_0)\}+
\alpha_2\Pr\{Z_{2,n} \in \Phi_i(R_0)\},
\]
from which together with (\ref{eq:donuts-11A}), (\ref{eq:donuts-12B})
it follows that 
$\nth\log\Pr\{Z_{n} \in \Phi_i(R_0)\}$ has the limit  such as 
\beqn\label{eq:donuts-14}
\lmn\nth\log\Pr\{Z_{n} \in \Phi_i(R_0)\} & =&
-\min\left(\inf_{R \in \Phi_i(R_0)}I_1(R),
\inf_{R \in \Phi_i(R_0)}I_2(R)\right).\nonumber\\
& &
\eeqn
As a consequence, from the definition of $\underline{H}(R)$, $\overline{H}(R)$ as in Section 
\ref{sec:fuji-dev-A-10s}, we have
\beqns
\underline{H}(R_0) & = & -\lim_{i\to\infty}
\lmn\nth\log\Pr\{Z_{n} \in \Phi_i(R_0)\}\\
 & = &\lim_{i\to\infty}\min\left(\inf_{R \in \Phi_i(R_0)}I_1(R),
\inf_{R \in \Phi_i(R_0)}I_2(R)\right)\\
& = &\min(I_1(R_0),I_2(R_0)),
\eeqns
where we have again invoked the closed convexity (and hence the continuity)
of  the functions $I_1(R), I_2(R)$ (cf. Rockafeller \cite{rock-book} ).
Therefore,
\beq\label{eq:donuts-15}
\underline{H}(R)=\min(I_1(R),I_2(R))\quad (\forall R\in\sR).
\eeq
Similarly, 
\beq\label{eq:donuts-16}
\overline{H}(R)=\min(I_1(R), I_2(R))\quad (\forall R\in\sR).
\eeq
These (\ref{eq:donuts-15}), (\ref{eq:donuts-16})  are the lower/upper IS rate functions of the arithmetic mean $\sZ=\{Z_n\}\noi$ for the  mixed source 
$\sX$. It should be remarked here that $I_(R)$, $I_2(R)$ are convex functions but 
$\min(I_1(R),I_2(R))$ is not necessarily convex, which means that the rate functions
$\underline{H}(R)$, $\overline{H}(R)$ are not necessarily convex. 
Also, it is easy to check that the source $\sZ=\{Z_n\}\noi$ is $E$-tight.
Thus, Corollary \ref{sing-Seoul1} together with  (\ref{eq:donuts-15}), (\ref{eq:donuts-16}) yields
the large deviation formula in the case of mixed sources:
\beqn\label{eq:donuts-24}
\lefteqn{-\inf_{R \in \Gamma^{\circ}}\min(I_1(R),I_2(R))\quad\quad}
\nonumber\\
 & \le &
\lin\nth\log\Pr\{Z_n \in \Gamma\} \nonumber\\
& \le &
\lsn\nth\log\Pr\{Z_n \in \Gamma\} 
\nonumber\\
\quad\quad & \le &
-\inf_{R \in \overline{\Gamma}}\min(I_1(R),I_2(R)). 
\eeqn
On the other hand, we recall that the cumulant generating function of 
$\sZ= \{Z_{n}\}\noi$ is given by 
\[
\varphi_n(\theta)=\nth \log \int P_{Z_n}(dz)e^{n\theta z},
\]
which is here written as 
\[
\varphi_n(\theta)=\nth \log \int \left(\alpha_1P_{Z_{1,n}}(dz)e^{n\theta z}
+\alpha_2P_{Z_{2,n}}(dz)e^{n\theta z}\right).
\]
It is not difficult to verify that  the limit
$\varphi(\theta)\equiv \lmn \varphi_n(\theta)$ exists with
\beq\label{eq:donuts-19}
\varphi(\theta)=
\max\left(\varphi_1(\theta), \varphi_2(\theta)\right).
\eeq
We notice  that the functions $\varphi(\theta), 
\varphi_1(\theta), \varphi_2(\theta)$ are always convex.
With this $\varphi(\theta)$  let us here define, as usual, the ``rate function" by 
\[
I(R)\equiv \sup_{\theta} (\theta R -\varphi(\theta)).
\]
However, this ``rate function" $I(R)$ is always convex and hence is different from
the lower/upper IS rate functions 
$\underline{H}(R), \overline{H}(R)$, because $\underline{H}(R), \overline{H}(R)$ are not necessarily convex.  Thus,  in the case of mixed sources $I(R)$ cannot be a  
``pertinent"  large deviation rate measure.

\bigskip

\noindent
{\em B.\ Nonstationary sources:}

\bigskip

Here too, we consider two Gaussian sources
\beq\label{eq:donuts-1m}
\sX_1 = (X_{1,1}, X_{1,2}, \cdots),
\eeq
\beq\label{eq:donuts-2l}
\sX_2 = (X_{2,1}, X_{2,2}, \cdots)
\eeq
 such as defined in (\ref{eq:donuts-1}) and (\ref{eq:donuts-2}).
Define the source $\sZ=\{Z_n\}\noi$ by
%
\beq\label{eq:osamuPPQb}
Z_n
 = \left\{ 
  \begin{array}{ccc}
 Z_{1,n}\equiv \nth\sum_{i=1}^{n}X_{1,i}
 & \mbox{ {\rm  if $n$ is odd,} }\\
 &\\
    Z_{2,n}\equiv\nth\sum_{i=1}^{n}X_{2,i}
 & \mbox{ {\rm  if $n$ is even.} }& 
  \end{array}
    \right.
\eeq 
Then, it is not difficult to verify that 
\beqn\label{eqbv:lanr5}
\overline{H}(R_0) & = & \lim_{i\to\infty}
\lsn \frac{1}{n}\log\frac{1}{\Pr\{Z_n \in \Phi_i(R_0\}}\nonumber\\
& =&  \max (\overline{H}_1(R_0), \overline{H}_2(R_0)),
\eeqn
where $\overline{H}(R_0)$, $\overline{H}_1(R_0), \overline{H}_2(R_0)$
are the upper IS rate functions for  
$\sZ=\{Z_{n}\}\noi, $ $\sZ_1=\{Z_{1,n}\}\noi,$ $ \sZ_2=\{Z_{2,n}\}\noi$, respectively.

Similarly,
 it is not difficult to verify also that 
\beqn\label{eqbv:lanrU}
\underline{H}(R_0) & = & \lim_{i\to\infty}
\lin \frac{1}{n}\log\frac{1}{\Pr\{Z_n \in \Phi_i(R_0\}}\nonumber\\
& =&  \min (\underline{H}_1(R_0), \underline{H}_2(R_0)),
\eeqn
where $\underline{H}(R_0)$, $\underline{H}_1(R_0), \underline{H}_2(R_0)$
are the lower IS rate functions for   
$\sZ=\{Z_{n}\}\noi, $ $\sZ_1=\{Z_{1,n}\}\noi,$ $ \sZ_2=\{Z_{2,n}\}\noi$, respectively.
Thus, we see that $\underline{H}(R_0)$ $\neq \overline{H}_1(R_0)$
in general.

Now, (\ref{eqbv:lanr5}) and (\ref{eqbv:lanrU}) are rewritten as
\beqn
\underline{H}(R_0) &=& \min (I_1(R_0), I_2(R_0)),\label{eqs:bsp1}\\
\overline{H}(R_0) &=& \max (I_1(R_0), I_2(R_0)),\label{eqs:bsp3}
\eeqn
because 
\beqns\label{eqbv:lanrUw}
\underline{H}_1(R_0) & = & \overline{H}_1(R_0) = I_1(R_0),\\
\underline{H}_2(R_0) & = & \overline{H}_2(R_0) = I_2(R_0).\
\eeqns
 Since it is easy to check that the source $\sZ=\{Z_n\}\noi$ is 
$E$-tight, $C$-tight and $\sigma$-convergent,  
Theorem \ref{teiri:spec-upper}
 and Theorem \ref{teiri:spec-lower}
 yield
\beqn\label{eq:hanTEFo1} 
\lefteqn{-\inf_{R \in \Gamma^{\circ} } \min( I_1(R), I_2(R))}\nonumber\\
&\le & \lsn\nth \log \Pr\{Z_n\in \Gamma\}\nonumber\\
&\le & -\inf_{R \in \overline{\Gamma} }\min( I_1(R), I_2(R)),
\eeqn 
and 
\beqn\label{eq:hanTEFo1q} 
\lefteqn{-\inf_{R \in \Gamma^{\circ} } \max I_1(R), I_2(R))}\nonumber\\
&\le & \lin\nth \log \Pr\{Z_n\in \Gamma\}\nonumber\\
&\le & -\inf_{R \in \overline{\Gamma} }\max( I_1(R), I_2(R)).
\eeqn 
Thus, in this nonstationary case, the large deviation principle cannot
be specified only with a {\em single} rate function
but can be with a {\em pair} of rate functions as in (\ref{eq:hanTEFo1} ) 
and (\ref{eq:hanTEFo1q}). Notice that Corollary \ref{sing-Seoul1} for this case does
not  work as well.
\section{Note on Generalizations\label{generalization-ex1}}
So far we have established two fundamental theorems 
(Theorem \ref{teiri:spec-upper}, Theorem \ref{teiri:spec-lower}) assuming that 
random variables $Z_n$ take values in the real space $\sR$. Actually, however,
we can generalize these theorems to the case where $Z_n$ takes values 
in a general topological space $\cX$. To see this, we extend Definitions
\ref{teigi:extra-1}, \ref{ISITA-S1}, \ref{ISITA-S2w}  as follows:
\bteigi\label{teigi:extra-1g}
\footnote{This definition is found in Dembo and Zeitouni \cite{demboS}} 
\beq\label{re;erq1}
\overline{H}(R) = \sup_{v(R)}\lsn \nth \log \frac{1}{\Pr\{Z_n \in v(R)\}},
\eeq
\beq\label{re;erq2}
\underline{H}(R) = \sup_{v(R)}\lin \nth \log \frac{1}{\Pr\{Z_n \in v(R)\}},
\eeq
where 
$\sup_{v(R)}$ denotes the supremum over all the neighborhoods 
$v(R)$ of $R$.\QED
\eteigi
\bteigi{\rm ($E$-tight)}\label{ISITA-S1g} \mbox{} 
If for any L there exists a compact set $\cA_L \subset \cX$ such that
\beq\label{eq:khan-italy1}
\lsn\nth\log \Pr\{Z_n \in \cA_{L}^c\} \le -L
\eeq
where $c$ indicates the complement of a set,
 then we say that the source $\sZ=$ $\{Z_n\}\noi$ is
 exponentially tight (abbreviated as $E$-tight;\ cf. Dembo and Zeitouni \cite{demboS}).
 \QED
\eteigi
\bteigi{\rm ($\sigma$-convergent)}\label{ISITA-S2w-gm}
Given a general source  $\sZ=$ $\{Z_n\}\noi$, set
\beq\label{eq:han1}
f_n(R) \equiv \frac{1}{n}\log \frac{1}{\Pr\{Z_n \in v(R)\}}\quad
 (\mbox{with any fixed neighborhood $v(R)$}).
\eeq
If  $\{f_n(R)\}\noi$ is $\sigma$-convergent in the sense of Definition \ref{ISITA-S2}, then we say that
the source $\sZ=$ $\{Z_n\}\noi$ is 
$\sigma$-convergent .\QED
\eteigi
\bchui\label{chui:saigo1}
As will be easily seen, Definitions \ref{teigi:extra-1g}, Definition \ref{ISITA-S1g},
Definition \ref{ISITA-S2w-gm}
reduce to Definition \ref{teigi:extra-1}, Definition \ref{ISITA-S1},
Definition \ref{ISITA-S2w}, respectively, in the case of  $\cX =\sR $.
  \QED
\echui
\bteiri\label{teiri:general-note} With these extended definitions, 
Therem \ref{teiri:spec-upper} and 
Theorem \ref{teiri:spec-lower} as well as Corollary \ref{sing-Seoul1}
and Corollary \ref{kei-hansi-God} hold  also  with 
 any  Hausdorff topological space $\cX$ instead of $\sR$. 
 \eteiri

\medskip

\noindent
{\em Proof}: \ It suffices basically to parallel the proofs of Theorem \ref{teiri:spec-upper} and 
Theorem \ref{teiri:spec-lower} with due modifications,
while paying attention to the fact that the intersection of a closed set $S$ and
a compact set $T$ is compact in the Hausdorff topological space $\cX$.
\QED
%
%
\section{Cumulant Generating Functions and Information-Spectrum Rate Functions
\label{ss:1seoulA}}  
Thus far having established the fundamental formulas ( Theorem \ref{teiri:spec-upper}, 
Theorem \ref{teiri:spec-lower}, Corollary \ref{sing-Seoul1}, 
Corollary \ref{kei-hansi-God}) on general large deviation problems described
in terms of the lower/upper IS rate functions 
$\underline{H}(R)$, $\overline{H}(R)$, we are now interested in
the problem of how to compute $\underline{H}(R)$, $\overline{H}(R)$
when the random variables $Z_n$ take values in $\sR$.

In many ``simple" source cases, for example, 
as Cram\'er's theorem and G\"artner-Ellis' theorem tell us, 
a desirable  large deviation ``rate function"  $I(R)$  is  computed 
as  the Fenchel-Legendre transforms
(cf. Rockafeller \cite{rock-book}) of the cumulant generatiing function 
 $\varphi (\theta)$ (or something like that):
\beq\label{eq:teikaku1d}
I(R) = \sup_{\theta}(\theta R - \varphi (\theta)).
\eeq
In such cases, the problem of computing the large deviation function $I(R)$ reduces to how to compute
the cumulant generating function  $\varphi (\theta)$. On the other hand, 
we notice here that, in the light of Theorem \ref{teiri:spec-upper} and   
Theorem \ref{teiri:spec-lower}, $\underline{H}(R)$ and $\overline{H}(R)$ also should 
be regarded as   ``rate functions," which suggests that  $\underline{H}(R)$, $\overline{H}(R)$ might be set to be equal to  $I(R)$  as:
\beq\label{eq:Akanori-1}
\underline{H}(R) = \overline{H}(R)=
\sup_{\theta}(\theta R - \varphi (\theta)).
\eeq

However, in most general cases that are not necessarily elementary and/or  typical, 
the right-hand side of (\ref{eq:teikaku1d}) does not give rise to a desirable  large deviation rate function any more, as was already seen in the foregoing section, i.e., 
 in general cases,\beq\label{eq:asahi-B}
\underline{H}(R), \overline{H}(R) \not =\sup_{\theta}(\theta R - \varphi (\theta)).
\eeq
Our main concern with (\ref{eq:Akanori-1}),  (\ref{eq:asahi-B}) then addresses the problem of  how to elucidate under what conditions  (\ref{eq:Akanori-1}) holds and/or 
under what conditions  (\ref{eq:Akanori-1}) does not hold; furthermore, also  if not then how not.

\medskip

In this section we address this  problem.
Let $\sZ =\{Z_n\}\noi$ be a general source.
To this end, let us start
with  any fixed closed interval $M\equiv [M_1, M_2] \ (M_1<M_2)$ 
and define the $M$-truncated cumulant generating function
by
\beq\label{eq:seoul-han1}
\varphi_n^{(M)}(\theta) \equiv\nth\log\int_{ M}P_{Z_n}(dz)e^{n\theta z},
\eeq
and also, define
\beq\label{eq:osamu-11}
\overline{\varphi}_M(\theta) \equiv \lsn \varphi^{(M)}_n(\theta)
\quad (\forall \theta \in \sR),
\eeq
\beq\label{eq:osamu-12}
\underline{\varphi}_M(\theta) \equiv \lin \varphi^{(M)}_n(\theta)
\quad (\forall \theta \in \sR).
\eeq
Moreover, define the $M$-truncated lower/upper 
IS rate functions
$\underline{H}_M(R)$,
 $\overline{H}_M(R)$ by
\beq\label{eq:osamu-9}
\underline{H}_M(R)
 = \left\{ 
  \begin{array}{ccc}
  \underline{H}(R) & \mbox{ for } &R\in M,\\
       +\infty                                 & \mbox{ for } & R \not\in M,
  \end{array}
    \right.
\eeq
\beq\label{eq:osamu-10}
\overline{H}_M(R)
 = \left\{ 
  \begin{array}{ccc}
  \overline{H}(R) & \mbox{ for } & R \in M,\\
       +\infty                                 & \mbox{ for } & R \not\in M,
  \end{array}
    \right.
\eeq
where $\underline{H}(R)$, $\overline{H}(R)$ are the lower/upper IS rate functions.
Then, we have the following fundamental formulas.
\bteiri\label{teiri:kihonasBan}
For any  general source $\sZ =\{Z_n\}\noi$ it holds that 
\beqn
\overline{\varphi}_M(\theta) & =&
\sup_{R}(\theta R-\underline{H}_M(R))
\quad(\forall \theta \in \sR).
\label{eq:osamu-20xxBAn}\\
\underline{\varphi}_M(\theta) & \ge  &                         
\sup_{R}(\theta R-\overline{H}_M(R))
\quad(\forall \theta \in \sR).
\label{eq:osamu-21xxBa}
\eeqn
\eteiri
The proof of Theorem \ref{teiri:kihonasBan} is given in Section \ref{setu:proofs}.\QED
%
%
\bchui\label{chui:seoul-han1mpBan}
In case $\sZ =\{Z_n\}\noi$ is   $\sigma$-convergent 
(cf. Definition \ref{ISITA-S2w})
in Theorem 
\ref{teiri:kihonasBan}, 
the following holds:
\beq
\underline{\varphi}_M(\theta) =
\sup_{R}(\theta R-\overline{H}_M(R))
\quad(\forall \theta \in \sR).
\label{eq:osamu-21-qkaBan}
\eeq
\echui

Let us now consider the special case  of 
(\ref{eq:seoul-han1})$\sim$ (\ref{eq:osamu-21-qkaBan})
 with $M=[-K, K]$ $(K>0)$. This special case is  indicated, with an abuse of notation, simply 
  by ``$K"$ in place of $``M."$
Then, an immediate consequence of Theorem \ref{teiri:kihonasBan} with 
$K$ in place of $M$  (under the limiting operation $K\to \infty$)  is the following theorem, where we have set

\beqn
\overline{\varphi}^{\circ}(\theta)\equiv   \lim_{K\to\infty}\overline{\varphi}_K(\theta) 
\quad(\forall \theta \in \sR),
\label{eq:osamu-20S}\\
\underline{\varphi}^{\circ}(\theta)\equiv   \lim_{K\to\infty}\underline{\varphi}_K(\theta)
\quad(\forall \theta \in \sR).
\label{eq:osamu-21T}
\eeqn
%
%
%
%
\bteiri\label{teiri:kihon}
For any  general source $\sZ =\{Z_n\}\noi$ it holds that
\beqn
\overline{\varphi}^{\circ}(\theta) & =&
\sup_{R}(\theta R-\underline{H}(R))
\quad(\forall \theta \in \sR),
\label{eq:osamu-20K}\\
\underline{\varphi}^{\circ}(\theta)& \ge&                            
\sup_{R}(\theta R-\overline{H}(R))
\quad(\forall \theta \in \sR).
\label{eq:osamu-21K}
\eeqn
\eteiri
\bchui\label{chui:seoul-han1}
In case  the  $\sigma$-convergence property is  satisfied in Theorem
\ref{teiri:kihon}, 
the following holds:
\beq
\underline{\varphi}^{\circ}(\theta) =
\sup_{R}(\theta R-\overline{H}(R))
\quad(\forall \theta \in \sR).
\label{eq:osamu-21-qa}
\eeq
\echui
{\em Proof of Theorem \ref{teiri:kihon}:}
\medskip

In view of Theorem \ref{teiri:kihonasBan}
 it suffices to take account of the definition of $\overline{\varphi}^{\circ}(\theta)$, $\underline{\varphi}^{\circ}(\theta)$
 and to notice that
\beqn
\lim_{K\to\infty}\sup_{R}(\theta R-\underline{H}_K(R))& =&
\sup_{R}(\theta R-\underline{H}(R))
\quad(\forall \theta \in \sR),\\
\lim_{K\to\infty}\sup_{R}(\theta R-\overline{H}_KR))
& =& \sup_{R}(\theta R-\overline{H}(R))   
\quad(\forall \theta \in \sR).                        
\label{eq:osamu-21xxBa2}
\eeqn
\QED

So far in Theorem \ref{teiri:kihonasBan} we have shown a relation between 
the $M$-truncated cumulant 
generating functions and the  $M$-truncated lower/upper IS rate functions.
We now want to see the direct ({\em not}  via $M$-truncation) relation between the non-truncated cumulant 
generating functions and the non-truncated lower/upper IS rate functions. 

The non-truncated cumulant generating functions  are defined by
\beq\label{eq:Seoul-Han2}
\varphi_n(\theta) \equiv\nth\log\int_{-\infty}^{+\infty} P_{Z_n}(dz)e^{n\theta z},
\eeq
and
\beq\label{eq:osamu-11a}
\overline{\varphi}(\theta) \equiv \lsn \varphi_n(\theta)
\quad (\forall \theta \in \sR),
\eeq
\beq\label{eq:osamu-12sa}
\underline{\varphi}(\theta) \equiv \lin \varphi_n(\theta)
\quad (\forall \theta \in \sR).
\eeq
Paralleling  the previous functions $\varphi_n^{(K)}(\theta), 
\varphi^{\circ}(\theta)$,
we  define the following ``tail" functions with an arbitrary $K>0$:
\beqn\label{eq:kanan-s1}
\varphi^{(\vee K)}_n(\theta) & \equiv  &\nth\log\int_{|z|>K}P_{Z_n}(dz)e^{n\theta z},\\
{\overline\varphi}^{\vee}(\theta) & \equiv& \lim_{K \to\infty}\lsn\varphi_n^{(\vee K)}
(\theta).\nonumber\label{eq:kanan-s2}
\eeqn
\bteigi{\rm( $C$-tight)}\label{eq:ttas} Let $\sZ=\{Z_n\}\noi$ be a general source.
If
\beq\label{eq:hamaleeq} 
\lim_{K \to\infty}\varphi^{(\vee K)}(\theta) = -\infty\quad (\forall\theta \in \sR),
\eeq
\eteigi
then we say that $\sZ=\{Z_n\}\noi$ is cumulatively tight
(abbreviated as $C$-tight).\QED
%

%
Then, we now have the following  lemma that relates the truncated cumulant generating functions to the non-truncated cumulant generating functions:
\bhodai\label{chui:seoul-han0ajk}
If a general source  $\sZ =\{Z_n\}\noi$
is  $C$-tight (cf. Definition \ref{ISITA-S1}),
 then it holds that
\beq\label{eq:bnmv1}
\overline{\varphi}^{\circ}(\theta)=\overline{\varphi}(\theta), \quad
\underline{\varphi}^{\circ}(\theta)=\underline{\varphi}(\theta)
 \quad ({\rm \forall\ } \theta \in \sR).\label{eq:bnmv2}
\eeq
The proof of this lemma is given in Section \ref{setu:proofs}. \QED
\ehodai

%
%
%
%

Now,  Lemma \ref{chui:seoul-han0ajk}, together with 
Theorem \ref{teiri:kihon} and  Remark \ref{chui:seoul-han1},
 immediately leads to 
the following Theorem \ref{teiri:kihonas} and  Remark \ref{chui:seoul-han1mp},
respectively.
\bteiri\label{teiri:kihonas}
If a general source $\sZ =\{Z_n\}\noi$ is $C$-tight, then
\beqn
\overline{\varphi}(\theta) & =&
\sup_{R}(\theta R-\underline{H}(R))
\quad\ ( \forall\theta \in \sR),
\label{eq:osamu-20xxas}
\\
\label{eq:osamu-20xxasari}
\underline{\varphi}(\theta)& \ge&                            
\sup_{R}(\theta R-\overline{H}(R))
 \quad ( \forall\theta \in \sR).
\label{eq:osamu-21xxwe}
\eeqn
\eteiri
\bchui\label{chui:seoul-han1mp}
In case $\sZ =\{Z_n\}\noi$ is $C$-tight and
 $\sigma$-convergent in Theorem 
\ref{teiri:kihonas}, 
the following holds:
\beq
\underline{\varphi}(\theta) =
\sup_{R}(\theta R-\overline{H}(R))
\quad (\forall\theta \in \sR).
\label{eq:osamu-21-qka}
\eeq
\echui
\bchui\label{chui:lancab1}
Theorem \ref{teiri:kihonas} along with Remark \ref{chui:seoul-han1mp}
is reminiscent of Varadhan's integral lemma \cite{varadhan} in which the linear function 
$\theta R$ of $R$ in the former is
replaced by an arbitrary  continuous function $\phi (R)$ on a regular topological space
$\cX$. However, the former does not follow as a special case from the latter,
because in the latter case
 the existence of a {\em good} rate function 
is assumed. \QED

\echui

Before proceeding to show  the IS formulas for rate functions
described in terms  of the cumulant generating functions, we need two definitions and one lemma.
\bteigi\label{teigi:roka-f} {\rm ({\rm Rockafeller \cite{rock-book}})} 
Given a function $f$  on {\bf R}, we define
the closed convex hull function $\sqcup  f$ of $f$ as 
the pointwise supremum of the collection of all 
affine functions $h$ on {\bf R} such that $h(R)$ $\le$ $f(R)$ ($\forall R \in \sR$).
It is evident that $\sqcup f(R)\le f(R)$ for all $R \in \sR$.
If  $\sqcup f(R)= f(R)$ for all $R \in \sR$, we say that 
$f$ is a closed convex function.
\eteigi
\bteigi\label{teigi:roka-weii}
{\rm ({\rm Fenchel-Legendre transform:\,Rockafeller \cite{rock-book}})} 
 If
\beq\label{eq:lpringer1}
g(\theta) =
\sup_{R}(\theta R-f(R))
\quad  (\forall \theta \in \sR),
\eeq
then we say that $g$ is the conjugate of $f$,
and denote the $g$ by $f^*$. 
\eteigi
\bhodai\label{hodai:hodaiQQ}{\rm ({\rm Rockafeller \cite{rock-book}})} 
The conjugate $f^*$ is always a closed convex function. Moreover,
it holds that $f^* = (\sqcup f)^*$ and
$f^{**} = \sqcup f.$\QED
\ehodai
%
 %
%

%
Thus,
applying Lemma \ref{hodai:hodaiQQ} to 
Theorem \ref{teiri:kihonas} and
Remark \ref{chui:seoul-han1mp}
 immediately yields the following inverse formulas:
\bteiri\label{teiri:kihonrrs}{\rm (Inverse formula)}
For any  $C$-tight source $\sZ =\{Z_n\}\noi$ it holds that
\beqn
\sqcup\underline{H}(R)& =&
\sup_{\theta}(\theta R-\overline{\varphi}(\theta))
\quad(\forall R \in \sR),
\label{eq:osamu-20Kqq}\\
\sqcup\overline{H}(R)
& \ge&                            
\sup_{\theta}(\theta R-\underline{\varphi}(\theta))
\quad(\forall R \in \sR).
\label{eq:osamu-21Krr}
\eeqn
\eteiri
\bchui\label{chui:seoul-han2}{\rm (Inverse formula)}
In case  the  $\sigma$-convergence property is  also satisfied in Theorem
\ref{teiri:kihonrrs}, 
the following holds:
\beq
\sqcup\overline{H}(R) =
\sup_{\theta}(\theta R-\underline{\varphi}(\theta))
\quad(\forall R \in \sR).
\label{eq:osamu-21-qazx}
\eeq
\echui
\bchui\label{chui:kondp10}
Theorem \ref{teiri:kihonrrs} as well as Remark \ref{chui:seoul-han2} is
reminiscent of Bryc's inverse Varadhan lemma \cite{bryc} 
 in which the  linear function $\theta R$ of $R$ in the former
 is replaced by an arbitray {\em bounded} continuous function $f(R)$
 on a completely regular topological space $\cX$. However, the former does not follow as a special case from the latter, because 
 linear functions are {\em not bounded}. 
 
 Furthermore, Theorem \ref{teiri:kihonrrs} as well as Remark \ref{chui:seoul-han2},
 together with Theorem \ref{teiri:kihonas} along with Remark \ref{chui:seoul-han1mp},  is
reminiscent of Dembo and Zeitouni \cite[Theorem 4.5.10]{demboS}, in which, however,
 the existence of a good rate function is assumed unlike in our case.
  \QED
\echui

A direct consequence of Theorem \ref{teiri:kihonas},
  Theorem \ref{teiri:kihonrrs},  Remark
\ref{chui:seoul-han1mp} and
 Remark \ref{chui:seoul-han2} is the following
 corollary which states a key relation 
 between the lower/upper IS rate
 functions $\underline{H}(R), \overline{H}(R)$ and the cumulant generating
 functions
$ \underline{\varphi}(\theta), \overline{\varphi}(\theta)$: 
\bkei{\rm (A condition for the normalized cumulant generating function
to have the limit)}\label{kei:france-1} \mbox{}

1)\ Let  a source $\sZ=\{Z_n\}\noi$ be $C$-tight. Then, 
\beq\label{eq:arirans-2}
\sqcup\underline{H}(R)=\sqcup\overline{H}(R)\quad
(\forall R \in \sR)
\eeq
 implies 
\beq\label{eq:arirans-3}
  \underline{\varphi}(\theta)=\overline{\varphi}(\theta)
\quad (\forall \theta \in \sR),
\eeq
  that is, 
 the cumulant generating function $\varphi_n(\theta)$
  defined by (\ref{eq:Seoul-Han2}) has the limit
\beq\label{eq:arirans-4}
\varphi(\theta) \equiv \underline{\varphi}(\theta)=\overline{\varphi}(\theta)
\quad (\forall \theta \in \sR).
\eeq

 \quad 2) If a source $\sZ=\{Z_n\}\noi$ is not only $C$-tight but also is 
 $\sigma$-convergent, then 
 (\ref{eq:arirans-2})
 is the necessary and sufficient condition for (\ref{eq:arirans-3}).
 \QED
 \ekei

\noindent
{\em Proof}: \ 1)\  Suppose that (\ref{eq:arirans-2})
holds. By Lemma \ref{hodai:hodaiQQ} combined with 
Theorem \ref{teiri:kihonas} we see that
\beqn
\overline{\varphi}(\theta) & =& \sqcup\overline{\varphi}(\theta) =
\sup_{R}(\theta R-\sqcup\underline{H}(R))
\quad\ {\rm for\ } \forall\theta \in \sR,
\label{eq:osamu-20xxasARI}
\\
\label{eq:osamu-20xxasAS1}
\underline{\varphi}(\theta) & \ge & \sqcup\underline{\varphi}(\theta)  \ge                            
\sup_{R}(\theta R-\sqcup\overline{H}(R))
 \quad {\rm for\ } \forall\theta \in \sR,
\label{eq:osamu-21xxweus}
\eeqn
which together with (\ref{eq:arirans-2}) yields 
$\overline{\varphi}(\theta)$ $=\underline{\varphi}(\theta)$ $(\forall \theta \in \sR)$,
where we have used the fact $\overline{\varphi}(\theta)$ $\ge
\underline{\varphi}(\theta)$.

\medskip

\noindent
\quad 2) \ Suppose that (\ref{eq:arirans-4}) holds. Then,  from 
(\ref{eq:osamu-20Kqq}) and (\ref{eq:osamu-21-qazx}) we have 
(\ref{eq:arirans-2}). \QED
%

\smallskip

In some sense, 
formulas (\ref{eq:osamu-20Kqq}) and (\ref{eq:osamu-21-qazx})
may be regarded as providing  formulas for computing the lower/upper IS 
rate functions $ \underline{H}(R), \overline{H}(R)$  as  the  Fenchel-Legendre transforms of  the cumulative generating functions 
 $\underline{\varphi}(\theta), \overline{\varphi}(\theta)$.
To see this more, let us define the following rate functions
$\underline{I}(R), \overline{I}(R)$ of
Cram\'er-G\"artner-Ellis type by
 \beq\label{ceq:stantanto2}
 \underline{I}(R)\equiv 
 \sup_{\theta}(\theta R -\overline{\varphi}(\theta))
\quad(\forall R \in \sR),
\eeq
 \beq\label{ceq:stantanto2Q}
 \overline{I}(R)\equiv 
 \sup_{\theta}(\theta R -\underline{\varphi}(\theta))
\quad(\forall R \in \sR).
\eeq
Then, from (\ref{eq:osamu-20Kqq}) and (\ref{eq:osamu-21-qazx})
we have 
\beq\label{ceq:stantanto3}
\sqcup\underline{H}(R) = \underline{I}(R),\quad
\sqcup\overline{H}(R)=\overline{I}(R)\quad (\forall R \in \sR).
\eeq
However,
in view  of Theorem \ref{teiri:spec-upper}, 
Theorem \ref{teiri:spec-lower} and Corollary \ref{sing-Seoul1}, 
Corollary \ref{kei-hansi-God},
 the formulas that  we wanted to obtain were those for computing 
$\underline{H}(R)$ =$\overline{H}(R)$ but not for
$\sqcup\underline{H}(R), \sqcup\overline{H}(R)$.
Formula (\ref{ceq:stantanto3}) tells us  that the ``rate function" 
$\underline{I}(R), \overline{I}(R)$ can
capture,  as well,  relevant  structures of large deviation probabilities 
that are  reflected  via the nature that   $\underline{I}(R), \overline{I}(R)$ are
 closed  convex functions (cf. Lemma \ref{hodai:hodaiQQ}). 
 In other words, $\underline{I}(R), \overline{I}(R)$ overlook all the finer structures that cannot be grasped
 via the closed convexity of $\underline{I}(R), \overline{I}(R)$ alone.
 We should be reminded that
 $\underline{H}(R), \overline{H}(R)$ are not necessarily  closed convex functions. 
 
 Thus, we do not yet reach the relevant formulas for computing 
 $\underline{H}(R),$ $\overline{H}(R)$ {\em via} the cumulant generating function,
 which remains to be further investigated.
 On the other hand, even without such relevant computation formulas,
 we could enjoy insightful general view,
 demonstrated so far in this paper, at basic large deviation problems.
 This is an advantage of the IS approach.
 
\medskip

These observations can formally be summarized as:
 
 %
 %
 
\bteiri\label{kei:river-ksks1}{\rm (Reduction theorem)}
Let  a general source
$\sZ=\{Z_n\}\noi$ be  $C$-tight and  $\sigma$-convergent.
Then,  it holds that 
\beq\label{eq:plusKAN}
\sqcup\underline{H}(R) = \underline{I}(R),\quad
\sqcup\overline{H}(R)=\overline{I}(R)\quad (\forall R \in \sR).
\eeq
In other words, 
it holds that
\beq\label{eq:plusKANsac}
\mbox{1)}\quad \underline{H}(R) =  \underline{I}(R) \quad(\forall R \in \sR)
\eeq
if and only if $\underline{H}(R)$ is closed and convex; and also that
\beq\label{eq:ogawa-print2}
\mbox{2)}\quad \overline{H}(R)=  \overline{I}(R) \quad(\forall R \in \sR)
\eeq
if  and only if $\overline{H}(R)$ is 
 closed and  convex  (cf. Definition \ref{teigi:roka-f}). 
\ ( Thus, in this case, the computation problems for $\underline{H}(R)$, 
 $\overline{H}(R)$ completely reduces  to those  for $\underline{I}(R)$, 
 $\overline{I}(R)$.)
\eteiri

\noindent
{\ Proof}:\quad The former part is the same one as in (\ref{ceq:stantanto3}).
 The latter part follows if we observe that 
$\underline{I}(R), \overline{I}(R)$ are always  closed convex functions.
 \QED
\bteiri\label{kei:river-ksks2}{\rm (General note)}
Let $\sZ=\{Z_n\}\noi$ be a general source and $\overline{\varphi}(\theta)$
be the cumulant generating function  defined by (\ref{eq:Seoul-Han2}), 
(\ref{eq:osamu-11a}). 
Then,
\beq\label{eq:suhyan-3}
\underline{H}(R)\ge 
\sqcup\underline{H}(R) \ge \underline{I}(R)\quad (\forall R \in \sR),
\eeq
where $\underline{I}(R)$ was defined in (\ref{ceq:stantanto2}).
\eteiri
\noindent
{\em Proof}:\quad The Fenchel-Legendre transformation of 
(\ref{eq:osamu-20K}) gives
\[
\sqcup\underline{H}(R) = \sup_{\theta}(\theta R - 
\overline{\varphi}^{\circ}(\theta)),
\]
from which together with $\overline{\varphi}(\theta)$  $\ge$
$\overline{\varphi}^{\circ}(\theta)$ it follows that
\[
\underline{H}(R) \ge \sqcup\underline{H}(R)\ge 
\sup_{\theta}(\theta R -\overline{\varphi}(\theta)) = \underline{I}(R).
\]
Thus, in this general case, the computation problem for $\underline{H}(R)$, 
 $\overline{H}(R)$ does {\em not} reduce to that for $\underline{I}(R)$.

\bchui\label{chui:ranranQRS2}
It is evident that $ \underline{I}(R)$ is a closed convex function.
Application of (\ref{eq:suhyan-3}) to
the right-most term in Theorem \ref{teiri:spec-upper}
yields 
a  Cram\'er-G\"artner-Ellis  type of upper bound (though in general much looser):
\beq\label{eq:yunijungr2}
\lsn \nth\log \Pr\{Z_n \in \Gamma\}
 \le -\inf_{R\in \Gamma}\underline{I}(R)
\eeq
for any compact set $\Gamma$ (cf. Dembo and Zeitouni \cite{demboS}). \QED
\echui
The following final remark concerns the ``locality" of the truncated Fenchel-Legendre transforms:
 \bchui{\rm (Locality)}\label{chui:Stanup-5}
 Let a source $\sZ=\{Z_n\}\noi$ be $C$-tight. We define 
$\underline{I}_M(R)$ as 
 \beq\label{eq:osamuPPQ}
 \underline{I}_M(R)
 = \left\{ 
  \begin{array}{ccc}
  \underline{I}(R) & \mbox{ {\rm for} } & R \in M,\\
       +\infty   & \mbox{ {\rm for} } & R \not \in M,
  \end{array}
    \right.
\eeq 
 and suppose that $ \underline{H}(R)$ is a closed convex function,
i.e., $\sqcup\underline{H}(R) =\underline{H}(R)$.
Then, from (\ref{eq:osamu-20Kqq}) we have $\underline{H}(R)=\underline{I}(R)$,
 and hence  $\underline{H}_M(R)$  $=\underline{I}_M(R)$
 $(\forall M, \forall R \in \sR).$  On the other hand, the he Fenchel-Legendre
  tansform of (\ref{eq:osamu-20xxBAn}) turns out to be
  \[
  \sqcup(\underline{H}_M)(R) = \sup_{\theta}(\theta R - \overline{\varphi}_M(R)).
  \]
Thus, in view of  $ \sqcup(\underline{H}_M)(R)=$ $\underline{H}_M(R)=$ 
$\underline{I}_M(R)$, we obtain

 \beq\label{qq:soeulqws1}
  \underline{I}_M(R)
 = \sup_{\theta}(\theta R - \overline{\varphi}_M(\theta)) \quad (\forall M=[M_1, M_2], \forall R \in \sR).
  \eeq
%
Then, (\ref{qq:soeulqws1}) means that, if we want to calculate the value of 
the rate function $\underline{I}(R)$ at some $R=a_0$, it is not necessary to calculate 
the values of 
$\overline{\varphi}(\theta)$ over all $\theta \in \sR$ and transform it. Instead, choose a small interval
$(c, d)$ contaning $a_0$ then it suffices 
to compute  the cumulant generating
function $\overline{\varphi}_M(\theta)$ only over the domain $M =[c, d]$
no matter how small it is.
This  demonstrates the ``locality" of the rate function $\underline{I}R)$.
Similarly for $\overline{I}(R)$, $\underline{\varphi}(\theta)$.
\QED
 \echui

\section{Proofs \label{setu:proofs}}
 
 In this section we give the proofs of Theorem \ref{teiri:kihonasBan}
 and Lemma \ref{chui:seoul-han0ajk}.
 \subsection{Proof of Theorem \ref{teiri:kihonasBan}\label{sub-big-5.1}}
 The proof of Theorem \ref{teiri:kihonasBan} consists of several steps,
 though they are elementary.
The mainstream is  to directly compute the cumulant generating function
with $M\equiv [M_1, M_2]\ (M_1<M_2)$:
\beq\label{eq:kanseikm}
\varphi_n^{(M)}(\theta)\equiv \nth\log \int_{M}P_{Z_n}(dz)e^{n\theta z}
\  \eeq
in terms of the quantities 
$\pi_1,\pi_2, \cdots$; $\Phi_i(R),$  $\overline{H}_i(R), \underline{H}_i(R)$,
$\overline{H}(R), \underline{H}(R)$
 defined as in the beginning of Section \ref{sec:fuji-dev-A-10s}.
\bigskip

\noindent
{\em Step 1}:

For each $i =1,2,\cdots$, set $\pi_i =2^{-i} $ and define an open interval
$\Phi_i(R)$ ( cf. Section \ref{sec:fuji-dev-A-10s}) by
\[
\Phi_i(R) = (R-\pi_i, R+\pi_i).
\]
Then, since $M=[M_1, M_2]$ is compact, for each $i$
there exists a finite number $L_i$ of  open intervals
\beq\label{eq:sojui1}
\Phi_i(a_i^{(j)})\quad (j=1,2, \cdots, L_i)
\eeq
such that
\beq\label{eqrq:ksnb}
a_i^{(j)} \in M \quad (j=1,2, \cdots, L_i)
\eeq
and 
\beq\label{eqrq:ksnbL}
 M \subset \bigcup_{j=1}^{L_i}\Phi_i(a_i^{(j)}).
\eeq
%
The collection (\ref{eq:sojui1}) of such open intervals is called a 
finite {\em cover} of $M$, simply denoted by $\Phi_i$.

Hereafter, for notational simplicity,  we write $I_i^{(j)}$ instead of $\Phi_i(a_i^{(j)})$.
Then, the integral of (\ref{eq:kanseikm}) is upper bounded as  
\beq\label{eq:isamu-8}
\nth\log \int_{M }P_{Z_n}(dz)e^{n\theta z}
\le \nth\log\sum_{j=1}^{L_i}\int_{I_i^{(j)}}P_{Z_n}(dz)e^{n\theta z}.
\eeq
%
On the other hand, by definition,
\beq\label{eqm:lanr1}
\underline{H}_i(a_i^{(j)})  =  \lin \frac{1}{n}\log\frac{1}{\Pr\{Z_n \in I_i^{(j)}\}},
\eeq
so that, 
for an arbitrarily small number $\delta >0$,
\beqn\label{eq:isamu-10}
\lefteqn{\Pr\{Z_n \in I_i^{(j)}\}}\nonumber\\
& \le &
\exp[-n(\underline{H}_i(a_i^{(j)})-\delta)]
\quad (\forall n\ge \exists n_i^{(j)};\ \forall j =1,2,\cdots, L_i).
\eeqn
Therefore，(\ref{eq:isamu-8}) is evaluated as follows.
\beqn\label{eq:isamu-11}
\int_{M}P_{Z_n}(dz)e^{n\theta z} 
&\le & \sum_{j=1}^{L_i}
\int_{ I^{(j)}_i }P_{Z_n}(dz)e^{n\theta z}\nonumber\\
& \le & \sum_{j=1}^{L_i}
\exp[-n(\underline{H}_i(a_i^{(j)}) -\delta)]
\exp[n\theta a_i^{(j)}+n\pi_i|\theta |]\nonumber\\
& = & \sum_{j=1}^{L_i}
\exp[-n(\underline{H}_i(a_i^{(j)}) -\delta)]
\exp[n\theta a_i^{(j)}+n2^{-i}|\theta |]\nonumber\\
&= &  \sum_{j=1}^{L_i}
\exp[n(\theta a_i^{(j)}-\underline{H}_i(a_i^{(j)}))]\exp[n(\delta +2^{-i}|\theta |)]
\nonumber\\
& \le & L_i\exp[ n\max_{j}(\theta a_i^{(j)}-\underline{H}_i(a_i^{(j)})) ]
\exp[n(\delta +2^{-i}|\theta |)].\nonumber\\
& &
\eeqn
Substituting (\ref{eq:isamu-11}) into the right-hand side (\ref{eq:isamu-8}) yields
\beq\label{eq:isamu-12}
\varphi_n^{(M)}(\theta) \le 
\max_{j}(\theta a^{(j)}_i-\underline{H}_i(a_i^{(j)})) +\delta +
2^{-i}|\theta | +\frac{1}{n}\log L_i.
\eeq
We now define the function $\underline{H}_i^{(M)}(R)$ on $\sR$ by
\beq\label{eq:isamu-12-hare-1}
\underline{H}_i^{(M)}(R)
 = \left\{ 
  \begin{array}{ccc}
 \underline{H}_i(R)& \mbox{ for } & R\in M,\\
       +\infty                                 & \mbox{ for } & R\not\in M.
  \end{array}
    \right.
\eeq
Then, (\ref{eq:isamu-12}) can be written as 
\beq\label{eq:isamu-13}
\varphi_n^{(M)}(\theta) \le 
\sup_{R}(\theta R-\underline{H}_i^{(M)}(R))
 +\delta + 2^{-i}|\theta |
++\frac{1}{n}\log L_i,
\eeq
where $\sup_{R}$ means the supremum over $\sR$.
Hence, 
\beq\label{eq:isamu-14}
\lsn \varphi_n^{(M)}(\theta) \le 
\sup_{R}(\theta R-\underline{H}_i^{(M)}(R)) +\delta + 2^{-i}|\theta |.
\eeq
It should be noted here that the function 
 $\underline{H}_i^{(M)}(R)$ is monotone increasing in $i$, that is，
\beq\label{eq:isamu-15}
\underline{H}^{(M)}_i (R) \le \underline{H}_{i+1}^{(M)}(R)
\quad (\forall R \in \sR; \  \forall i=1,2,\cdots).
\eeq
Therefore, we have the limit function (as was already defined by 
(\ref{eq:osamu-9})):
\beq\label{eq:isamu-16}
\underline{H}_M(R) = \lim_{i\to\infty}\underline{H}^{(M)}_i(R) \quad (\forall R \in \sR),
\eeq
where the value $+\infty$ is also allowed．
Now，by (\ref{eq:isamu-14}),
\beqn\label{eq:isamu-17}
\lsn \varphi_n^{(M)}(\theta)  & \le &
\lim_{i\to\infty}\left(\sup_{R}(\theta R-\underline{H}^{(M)}_i(R) +\delta + 2^{-i}|\theta |\right)\nonumber\\
  & \le &
\lim_{i\to\infty}\sup_{R}(\theta R-\underline{H}^{(M)}_i(R)) +\delta.
\eeqn
Since $\delta >0$ is arbitrarily small，it follows from (\ref{eq:isamu-17}) that
\beqn\label{eq:isamu-18}
\overline{\varphi}_M(\theta) \equiv \lsn \varphi_n^{(M)}(\theta) 
& \le &
 \lim_{i\to\infty}\sup_{R}(\theta R-\underline{H}^{(M)}_i(R))\nonumber\\
&= &
 \lim_{i\to\infty}\sup_{R}(\theta R-\sqcup\underline{H}^{(M)}_i(R)),
\eeqn
where,  for a function $f(R)$ on $\sR$, the function $\sqcup f$
is defined in Definition \ref{teigi:roka-f} (also see Lemma \ref{hodai:hodaiQQ}). 

\bigskip

\noindent
{\em Step 2}:

Define
\beq\label{eq:boston-232-1}
g_{\theta}^{(i)}(R) = \theta R -\sqcup\underline{H}_i^{(M)}(R)
\eeq
and set
\beq\label{eq:boston-232-2}
g_{\theta}^{(i)} = \sup_Rg_{\theta}^{(i)}(R),
\eeq
\beq\label{eq:boston-232-3}
g_{\theta}=\lim_{i\to\infty}\sup_Rg_{\theta}^{(i)}(R).
\eeq
Suppose here that $g_{\theta}=-\infty$, then it trivially holds that
\beqn\label{eq:boston-232-4}
\lim_{i\to\infty}\sup_R
\left(\theta R -\underline{H}_i^{(M)}(R)\right)
& =& \lim_{i\to\infty}\sup_R
\left(\theta R -\sqcup\underline{H}_i^{(M)}(R)\right)
\nonumber\\
& \le & \sup_R
\left(\theta R -\underline{H}_M(R)\right). 
\eeqn
%
Next, let us consider  the case of $g_{\theta}>-\infty$
and define the set:
\[
\cG_{\theta}(i)
 = \left\{R \in \sR \left|g_{\theta}^{(i)}(R)\ge g_{\theta}\right.\right\}.
\]
Then, we see that 
$\sup_R$ on the right-hand side of (\ref{eq:boston-232-2}) is attained at some point $R=R_0 \in M$; and  $g_{\theta}^{(i)}(R)$ is a closed concave function; so that  
$\cG_{\theta}(i)$ $(i=1,2,\cdots)$ is a sequence of monotone shrinking closed intervals.
 Therefore, there must exist 
at least a point $R_1 \in \sR$ such that 
\[
R_1 \in \bigcap_{i=1}^{\infty}\cG_{\theta}(i)
\]
and
\beq\label{eq:boston-232-5}
\theta R_1 -\sqcup\underline{H}_M(R_1)
= \lim_{i\to\infty}
\left(\theta R_1 -\sqcup\underline{H}_i^{(M)}(R_1)\right)\ge g_{\theta},
\eeq
where we have used  the monotonicity of 
$\sqcup\underline{H}_i^{(M)}(R)$ in $i$: 
\[
\lim_{i\to\infty}\sqcup\underline{H}_i^{(M)}(R) = 
\sqcup\underline{H}_M(R)\quad (\forall R \in \sR)
\]
Thus, 
from (\ref{eq:boston-232-1})$\sim$(\ref{eq:boston-232-3})
and (\ref{eq:boston-232-5}), we have
\beq\label{eq:boston-232-6}
\lim_{i\to\infty}\sup_R
\left(\theta R -\sqcup\underline{H}_i^{(M)}(R)\right)
\le \sup_R\left(\theta R -\sqcup\underline{H}_M(R)\right).
\eeq
Then, taking account of
\[
\sup_R\left(\theta R -\sqcup\underline{H}_i^{(M)}(R)\right) = 
\sup_R\left(\theta R -\underline{H}_i^{(M)}(R)\right),
\]
\[
\sup_R\left(\theta R -\sqcup\underline{H}_M(R)\right) = 
\sup_R\left(\theta R -\underline{H}_M(R)\right),
\]
we see that  (\ref{eq:boston-232-6}) is equivalent to 
\beq\label{eq:boston-232-7}
\lim_{i\to\infty}\sup_R
\left(\theta R -\underline{H}_i^{(M)}(R)\right)
\le \sup_R\left(\theta R -\underline{H}_M(R)\right).
\eeq
Consequently，by means of (\ref{eq:isamu-18}) and  (\ref{eq:boston-232-7}), it is concluded that 
\beq\label{eq:boston-232-8}
\overline{\varphi}_M(\theta) \le \sup_R\left(\theta R -\underline{H}_M(R)\right).
\eeq

\medskip

\noindent
{\em Step 3}:

Next, it follows from the definition of $\underline{H}_i(R)$ as in (\ref{seoul-sita13})
that, for any small $\delta>0$, there exists a sequence of positive integers
$n_1^{(j)}<n_2^{(j)}<\cdots \to\infty$, which may depend on $\delta>0$ and $i$,
such that 
\beqn\label{eq:isamu-18-app-1}
\lefteqn{\Pr\{Z_{n_k^{(j)}} \in I^{(j)}_i\}  \ge 
\exp[-n_k^{(j)}(\underline{H}_i(a_i^{(j)}) +\delta)]}\nonumber\\
& &\quad\quad\quad
 (\forall k\ge k_0(i,\delta);\ \forall i \ge i_0(\delta);\ \forall j =1,2,\cdots, L_i).
\eeqn
Then, 
\beqn\label{eq:isamu-18-app-2}
\lefteqn{\max_j\left(\frac{1}{n_k^{(j)}}
\log \int_{M}
P_{Z_{n_k^{(j)}}}(dz)e^{n_k^{(j)}\theta z}\right) }\nonumber\\
& \ge & 
\max_j\left(\frac{1}{n_k^{(j)}} 
\log \int_{ I^{(j)}_i}P_{Z_{n_k^{(j)}}}(dz)e^{n_k^{(j)}\theta z}
\right) \nonumber\\
& \ge & \max_j\left(\frac{1}{n_k^{(j)}} \log \left(
\exp[-{n_k^{(j)}}(\underline{H}_i(a_i^{(j)}) +\delta)]
\exp[{n_k^{(j)}}\theta a^{(j)}_i-n_k^{(j)}\pi_i|\theta |]\right)\right)
\nonumber\\
& = & \max_j\left(\frac{1}{n_k^{(j)}} \log \left(
\exp[-{n_k^{(j)}}(\underline{H}_i(a_i^{(j)})+\delta)]
\exp[{n_k^{(j)}}\theta a^{(j)}_i -n_k^{(j)}2^{-i}|\theta |]\right)\right)
\nonumber\\
&= & \max_j\left(
(\theta a^{(j)}_i-\underline{H}_i(a_i^{(j)}) )
-(\delta +2^{-i}|\theta |)\right)
\nonumber\\
& = & \max_{j}(\theta a^{(j)}_i-\underline{H}_i(a_i^{(j)}))
-(\delta +2^{-i}|\theta |).\nonumber\\
& &
\eeqn
As a consequence,
\beq\label{eq:isamu-18-app-3}
\max_j \varphi_{n_k^{(j)}}^{(M)}(\theta) \ge 
\max_{j}(\theta a^{(j)}_i-\underline{H}_i(a_i^{(j)})) -\delta -
2^{-i}|\theta |.
\eeq
Set
\[
\varphi_{n_k}^{(M)}(\theta) = \max_j \varphi_{n_k^{(j)}}^{(M)}(\theta)
\quad(n_k = n_k^{(j)};\ \exists j=1,2,\cdots, L_i),
\]
then (\ref{eq:isamu-18-app-3}) yields
\beq\label{eq:petersburg-1}
\varphi_{n_k}^{(M)}(\theta) \ge 
\max_{j}(\theta a^{(j)}_i-\underline{H}_i(a_i^{(j)})) -\delta -
2^{-i}|\theta |.
\eeq
We observe here that the left-hand side of (\ref{eq:petersburg-1})
does not depend on the choice of a finite cover $\Phi_i$ of $M$, so that
\beq\label{eq:petersburg-kq1}
\varphi_{n_k}^{(M)}(\theta) \ge \sup_{\Phi_i}
\max_{j}(\theta a^{(j)}_i-\underline{H}_i(a_i^{(j)})) -\delta -
2^{-i}|\theta |,
\eeq
where $\sup_{\Phi_i}$ means the supremum over all the finite covers 
${\Phi_i}$ of $M$.
It is easy to check that 
\[
\sup_{\Phi_i}
\max_{j}(\theta a^{(j)}_i-\underline{H}_i(a_i^{(j)}))
=\sup_{R}(\theta R-\underline{H}_i^{(M)}(R)).
\]
Thus, 
\beq\label{eq:isamu-18-app-3-1}
\varphi_{n_k}^{(M)}(\theta) \ge 
\sup_{R}(\theta R-\underline{H}_i^{(M)}(R))
 -\delta - 2^{-i}|\theta |. 
\eeq
Noting  that $n_k^{(j)}\ge k$ $(\forall j = 1,2, \cdots, L_i)$ and 
taking $\limsup_{k\to\infty}$ in both sides of
(\ref{eq:isamu-18-app-3-1}), we have
\beqn\label{eq:isamu-18-app-4}
\lefteqn{\overline{\varphi}_M(\theta) \equiv \lsn \varphi_n^{(M)}(\theta)}
 \nonumber\\& \ge &
\limsup_{k\to\infty} \varphi_{n_k}^{(M)}(\theta) \nonumber\\
& \ge &
\sup_{R}
(\theta R-\underline{H}^{(M)}_i(R))-\delta - 2^{-i}|\theta |
\nonumber\\
& \ge &
\sup_{R}
(\theta R-\underline{H}_M(R))-\delta - 2^{-i}|\theta |,
\eeqn
where we have taken account of the monotonicity of $\underline{H}_i^{(M)}(R)$.
Moreover, taking account of $\lim_{i\to\infty}$ in both sides of 
(\ref{eq:isamu-18-app-4}) and recalling that  $\delta>0$ is arbitrary, we have
\beq\label{eq:isamu-18-app-5}
\overline{\varphi}_M(\theta) \ge 
 \sup_{R}(\theta R-\underline{H}_M(R)).
\eeq
Then, it follows from (\ref{eq:boston-232-8}) and (\ref{eq:isamu-18-app-5})
that
\beq\label{eq:isamu-18-app-6}
\overline{\varphi}_M(\theta) \equiv \lsn \varphi_n^{(M)}(\theta)=
 \sup_{R}(\theta R-\underline{H}_{M}(R)) \quad (\forall \theta \in \sR),
\eeq
which is nothing but (\ref{eq:osamu-20xxBAn}). 

\bigskip

\noindent
{\em Step 4}:

Accordingly to (\ref{eqm:lanr1}), we set
\beq\label{eqm:TAnk2}
\overline{H}_i(a_i^{(j)}) =  
\lsn \frac{1}{n}\log\frac{1}{\Pr\{Z_n \in I_i^{(j)}\}}
\eeq
and define the function $\overline{H}_i^{(M)}(R)$ on $\sR$ as
\beq\label{eq:isamu-20}%
\overline{H}_i^{(M)}(R)
 = \left\{ 
  \begin{array}{ccc}
  \overline{H}_i(R) & \mbox{ for } & R\in M,\\
       +\infty                                 & \mbox{ for } & R\not\in M.
         \end{array}
    \right.
\eeq
Taking account of the monotonicity in $i$ of  
$\overline{H}_i^{(M)}(R) $, we obtain the limit function
\beq\label{eq:isamu-21}
\overline{H}_M(R) = \lim_{i\to\infty}\overline{H}_i^{(M)}(R) \quad (\forall R \in \sR).
\eeq
Then, in   the same manner as  in deriving (\ref{eq:isamu-18-app-5}),
we have
\beq\label{eq:bikkuri-1}
\underline{\varphi}_M(\theta) \equiv \lin \varphi_n^{(M)}(\theta) \ge 
 \sup_{R}(\theta R-\overline{H}_M(R)),
\eeq
which is nothing but (\ref{eq:osamu-21xxBa}).

\bigskip

\noindent
{\em Step 5}:

On the other hand, the opposite inequality:
\beq\label{eq:bikkuri-2}
\underline{\varphi}_M(\theta) \equiv \lin \varphi_n^{(M)}(\theta) \le 
 \sup_{R}(\theta R-\overline{H}_M(R))
\eeq
does {\em not } necessarily hold.
In the sequel we will show that the assumed $\sigma$-convergence on the interval 
$\cD=M$
(cf. Definitions \ref{ISITA-S2}, \ref{ISITA-S2w}) is a sufficient condition for 
(\ref{eq:bikkuri-2}) to hold.

In view of (\ref{eqm:TAnk2}) we see that
  the $\sigma$-convergence ( with any small $\delta$)
means the existence of 
a sequence  $n_1<n_2<\cdots\to\infty$, which may depend on 
$i$ and $ \delta$ but must not on $j$,  such that 
\beqn\label{eq:bikkuri-3}
\lefteqn{\frac{1}{n_k}\log \frac{1}{\Pr\{Z_{n_k} \in I^{(j)}_i\}}
 \ge  \overline{H}_i(a_i^{(j)})  -\delta}
\nonumber\\
& & \quad \quad
(\forall k\ge k_0(i,j,\delta);\  \forall i  \ge i_0(\delta);\ \forall j= 1,2,\cdots, L_i).
\eeqn
We  rewrite this as
\beqn\label{eq:bikkuri-4}
\lefteqn{\Pr\{Z_{n_k} \in I^{(j)}_i\} \le 
\exp[-n_k( \overline{H}_i(a_i^{(j)}) -\delta)]}\nonumber\\
& &\quad\quad\quad 
(\forall k\ge k_0(i,j,\delta);\  \forall i \ge i_0(\delta);\  \forall j= 1,2,\cdots, L_i).
\eeqn
Then, 
\beqn\label{eq:bikkuri-6}
\lefteqn{\int_{ I^{(j)}_i}P_{Z_{n_k}}(dz)e^{n_k\theta z} }\nonumber\\
& \le &
\exp[-n_k(\overline{H}_i(a_i^{(j)}) -\delta)]
\exp[n_k\theta a^{(j)}_i+n_k2^{-i}|\theta |]\nonumber\\
&= & 
\exp[n_k(\theta a^{(j)}_i-\overline{H}_i(a_i^{(j)}) )]
\exp[n_k(\delta +2^{-i}|\theta |)].
\nonumber\\
& &
\eeqn
Consequently,
\beqn\label{eq:bikkuri-7}
\lefteqn{\int_{M}P_{Z_{n_k}}(dz)e^{n_k\theta z} }\nonumber\\
&\le& \sum_{j=1}^{L_i}
\int_{ I^{(j)}_i}P_{Z_{n_k}}(dz)e^{n_k\theta z} \nonumber\\
& \le &\sum_{j=1}^{L_i}
\exp[n_k(\theta a^{(j)}_i-\overline{H}_i(a_i^{(j)}))]
\exp[n_k(\delta +2^{-i}|\theta |)]\nonumber\\
&\le &  L_i\exp[n_k\max_j (\theta a^{(j)}_i-\overline{H}_i(a_i^{(j)}) )]
\exp[n_k(\delta +2^{-i}|\theta |)]
\nonumber\\
& &
\eeqn
Therefore，
\beqn\label{eq:bikkuri-8}
\varphi_{n_k}^{(M)}(\theta) & \equiv & 
\frac{1}{n_k}\log \int_{M}P_{Z_{n_k}}(dz)e^{n_k\theta z} 
\nonumber\\
& \le &\max_{j}(\theta a^{(j)}_i-\underline{H}_i(a_i^{(j)})
+\delta +
2^{-i}|\theta | +\frac{\log L_i}{n_k}\nonumber\\
& \le & \sup_R(\theta R -\overline{H}_i^{(M)}(R))
+\delta +
2^{-i}|\theta | +\frac{\log L_i}{n_k}.
\nonumber\\
& &
\eeqn
Hence,
\beqn\label{eq:tamagawa-3}
\underline{\varphi}_M(\theta) & \equiv & \lin \varphi_n^{(M)}(\theta)\nonumber\\
& \le & \liminf_{k\to\infty}\varphi_{n_k}^{(M)}(\theta)\nonumber\\
& = & \sup_R(\theta R -\overline{H}_i^{(M)}(R))
+\delta +
2^{-i}|\theta |.
\eeqn
Now, taking $\lim_{i\to\infty}$ and letting $\delta\to 0$ in both of 
(\ref{eq:tamagawa-3}), we have
\[
\underline{\varphi}_M(\theta) \equiv \lin \varphi_n^{(M)}(\theta)
\le \lim_{i\to\infty}\sup_R(\theta R -\overline{H}_i^{(M)}(R)).
\]
Then, in an analogous manner as in   the argument
for a sequence of monotone shrinking intervals (Step 2),
we conclude that
\beq\label{eq:mantomanq}
\underline{\varphi}_M(\theta)  \le \sup_{R}(\theta R-\overline{H}_M(R))
  \quad (\forall \theta \in \sR),
\eeq
which together with (\ref{eq:bikkuri-1}) yields 
\beq\label{eq:isamu-22}
\underline{\varphi}_M(\theta) \equiv \lin \varphi_n^{(M)}(\theta) = \sup_{R}(\theta R-\overline{H}_M(R))  \quad (\forall \theta \in \sR).
\eeq

 \medskip
 \subsection{Proof of Lemma \ref{chui:seoul-han0ajk}\label{sub-big-5.2}} 
 \medskip

Since
\beqns
\lefteqn{\varphi_n(\theta) = \nth\log\int P_{Z_n}(dz)e^{n\theta z}}\\
& = &
 \nth\log\left(\int_{|z|\le K} P_{Z_n}(dz)e^{n\theta z}
+\int_{|z|>K} P_{Z_n}(dz)e^{n\theta z}\right)\\
&\le &  \nth\log\max\left(\int_{|z|\le K} P_{Z_n}(dz)e^{n\theta z},
\int_{|z|>K} P_{Z_n}(dz)e^{n\theta z}\right)+\nth\log 2\\
&=&\max\left( \nth\log\int_{|z|\le K} P_{Z_n}(dz)e^{n\theta z},
 \nth\log\int_{|z|>K} P_{Z_n}(dz)e^{n\theta z}\right)+\nth\log 2\\
&= & \max\left(\varphi^{(K)}_n(\theta), \varphi^{(\vee K)}_n(\theta) \right)
+\nth\log 2,
\eeqns
we have
\beqn\label{eq:cosmosp-3}
\overline{\varphi}(\theta)　& \equiv &\lsn \varphi_n(\theta)\nonumber \\
&= & \lsn \nth\log\int P_{Z_n}(dz)e^{n\theta z}\nonumber\\
&\le & \max \left( \lsn\varphi^{(K)}_n(\theta),
 \lsn \varphi_n^{(\vee K)}(\theta)\right).
\eeqn
 Letting $K$ $\to\infty$ in both sides of (\ref{eq:cosmosp-3}), 
it follows from the assumed $C$-tightness that 
\beqn\label{eq:cosmosp-4}
\overline{\varphi}(\theta)
& \le & \max\left( \lim_{K\to\infty}\lsn \varphi_n^{(K)}(\theta),
 \lim_{K\to\infty}\lsn \varphi_n^{(\vee K)}(\theta)\right)\nonumber\\
& = & \max\left(\overline{\varphi}^{\circ}(\theta), -\infty\right)\nonumber\\
& = & \overline{\varphi}^{\circ}(\theta).
\eeqn
On the other hand,  $\varphi(\theta) \ge \varphi^{\circ}(\theta)$
always holds, so that we conclude \beq\label{eq:boston-f9}
\overline{\varphi}(\theta) 
= \overline{\varphi}^{\circ}(\theta) \quad (\forall \theta \in \sR).
\eeq
Furthermore, in the same way as above,
we have
\beqn\label{eq:cosmosp-5}
\underline{\varphi}(\theta)& \equiv &\lin \varphi_n(\theta)\nonumber \\
&\le & \max
\left( \lin \varphi^{(K)}_n(\theta), \lsn \varphi_n^{(\vee K)}(\theta)\right).
\eeqn
 Again, letting $K\to\infty$ in (\ref{eq:cosmosp-5}) yields
\beqn\label{eq:cosmosp-6}
\underline{\varphi}(\theta)
&\le & \max\left( \lim_{K\to\infty}\lin \varphi_n^{(K)}(\theta),
\lim_{K\to\infty}\lsn \varphi_n^{(\vee K)}(\theta)\right)\nonumber\\
&=&\max \left(\underline{\varphi}^{\circ}(\theta), -\infty\right)\nonumber\\
& =& \underline{\varphi}^{\circ}(\theta),
\eeqn
where we have used again the assumed $C$-tightness.
Since $\underline{\varphi}(\theta) \ge \underline{\varphi}^{\circ}(\theta)$
always holds, we conclude that
\beq\label{eq:boston-f11}
\underline{\varphi}(\theta)  = \underline{\varphi}^{\circ}(\theta) 
\quad (\forall \theta \in \sR).
\eeq
%

%
\end{document}